\documentclass{aa}  

\usepackage{graphicx}
\usepackage{natbib}
\bibpunct{(}{)}{;}{a}{}{,} 
\usepackage{txfonts}
\usepackage{color,soul}

\newcommand\be{\begin{equation}}
\newcommand\en{\end{equation}}
\newcommand{\msun}{\mbox{\rm $M_{\odot}$}}

\newcommand{\lsun}{\mbox{\rm $L_{\odot}$}}

\newcommand{\jhk}{\mbox{$JHK_{\rm s}$}~}
\newcommand{\ks}{\mbox{$K_{\rm s}$}}

\newcommand{\av}{\mbox{$A_{\rm V}$~}}

\newcommand{\rv}{\mbox{$R_{\rm V}$~}}
\newcommand{\hii}{H\,{\footnotesize II}~}

\newcommand\cmsq{cm$^{-2}$~}

\begin{document}

\title{Deep GeMS/GSAOI Near-Infrared observations of N159W in the Large
  Magellanic Cloud} \subtitle{}

\author{ A. Bernard \inst{1,2} \and B. Neichel \inst{1} \and M. R.
  Samal\inst{1} \and A. Zavagno\inst{1} \and M. Andersen\inst{3} \and
  C. J. Evans\inst{4} \and H. Plana \inst{5} \and T. Fusco\inst{1,2} }
\titlerunning{Deep GeMS/GSAOI near-IR observations of N159W in the LMC}

  \institute{Aix Marseille Universit\'e, CNRS, LAM (Laboratoire
  d'Astrophysique de Marseille) UMR 7326, 13388, Marseille, France \\
  \email{anais.bernard@lam.fr} 
  \and 
  ONERA (Office National d'Etudes et de Recherches A\'erospatiales), B.P.72, F-92322 Chatillon, France
  \and 
  Gemini Observatory, c/o AURA, Casilla 603, La Serena, Chile
  \and 
  UK ATC, Royal Observatory, Blackford Hill, Edinburgh, EH9 3HJ, UK 
  \and Laboratorio de Astrof\'isica Te\'orica e Observacional,
  Universidade Estadual de Santa Cruz, Rodovia Jorge Amado km16
  45662-900 Ilh\'eus BA, Brazil}
  
   \date{}

 
  \abstract
  {}
  {The formation and properties of star clusters at the edge of
    H\,{\scriptsize II} regions are poorly known, partly due to
    limitations in angular resolution and sensitivity, which become
    particularly critical when dealing with extragalactic clusters. In
    this paper we study the stellar content and star-formation
    processes in the young N159W region in the Large Magellanic
    Cloud.}
  {We investigate the star-forming sites in N159W at unprecedented
    spatial resolution using {\mbox{$JHK_{\rm s}$}}-band images
    obtained with the GeMS/GSAOI instrument on the Gemini South
    telescope. The typical angular resolution of the images is
    $\sim$100\,mas, with a limiting magnitude of $H$\,$\sim$22\,mag
    (90\% completeness).  Photometry from our images is used to
    identify candidate young stellar objects (YSOs) in N159W. We also
    determine the $H$-band luminosity function of the star cluster at
    the centre of the H\,{\scriptsize II} region and use this to
    estimate its initial mass function (IMF).}
  {We estimate an age of 2\,$\pm$\,1\,Myr for the central cluster,
    with its IMF described by a power-law with an index of
    $\Gamma$\,$=$\,$-$1.05\,$\pm$\,0.2, and with a total estimated
    mass of $\sim$1300\,\msun. We also identify 104 candidate YSOs,
    which are concentrated in clumps and subclusters of stars,
    principally at the edges of the H\,{\scriptsize II} region. These
    clusters display signs of recent and active star-formation such as
    ultra-compact H\,{\scriptsize II} regions, and molecular outflows.
    This suggests that the YSOs are typically younger than the central
    cluster, pointing to sequential star-formation in N159W, which has
    probably been influenced by interactions with the expanding
    H\,{\scriptsize II} bubble.}
   {}


   \maketitle
%

\section{Introduction}
\label{sec:intro}

Feedback from massive stars ($M$\,$>$\,8\,\msun) plays a major role in
shaping the appearance of their surroundings. Their strong ultraviolet
radiation fields ionise the local interstellar medium (ISM), and their
intense stellar winds and eventual supernova explosions drive the dynamics
of the ISM, while also enriching it with heavy elements from the products
of nuclear fusion. 
However, despite their dominant role in galactic evolution, the
formation of massive stars is still not well understood.

Two main models have been proposed to explain the formation of massive
stars: monolithic collapse \citep{McKee2003} or competitive accretion
\citep{Bonnell2001}\footnote{See \citet{Tan2014} and
  \citet{Krumholz2015} for recent reviews.}. In the first, the
formation scenario is similar to that in low-mass stars but with high
accretion rates (and accretion disks), while in the latter the
protostars grow in mass from a surrounding gas reservoir (via
filamentary feeding) depending on their position in the clump. To
understand which model and/or process is the main agent in the
formation of massive stars, it is important to observe them, and their
host clusters, in their earliest phases. However, due to the nature of
the stellar initial mass function \citep[IMF;][]{Salpeter1955},
massive stars are inherently rare objects.

The earliest phases of massive stars evolve on short timescales
($\sim$10$^5$\,yr) and occur deeply embedded in their natal clumps,
making them difficult to observe \citep{Churchwell2002, Lada2003}. In
this context, recent wide-area, near-infrared (NIR) surveys with the
UKIRT and VISTA telescopes \citep[e.g.][]{Lawrence2007,
  Minniti2010,Cioni2011} have enabled detections of numerous candidate
star clusters and star-forming regions
\citep[e.g.][]{Borissova2011,Piatti2014,Romita2016}. But
characterising the stellar content of such clusters remains an
observational challenge, mainly due to limitations in angular
resolution and sensitivity, which become even more critical in
extragalactic systems.

The Large Magellanic Cloud (LMC) contains a rich sample of molecular
clouds, and its distance
\citep[$\sim$50\,kpc, e.g.][]{Pietrzynski2013,deGrijs2014} is small
enough to allow study of individual objects within those clouds.  The
LMC is also sufficiently far from the Galactic plane
($b$\,$\sim$$-$33$^\circ$) that foreground extinction is relatively
small. As such, the LMC provides an excellent system in which to study
massive star-formation regions in an extragalactic environment
\citep[with a metallicity about half of the solar value,
e.g.][]{Dufour1982,Smith1999}

\begin{figure}
   \centering
 \includegraphics[width=\hsize]{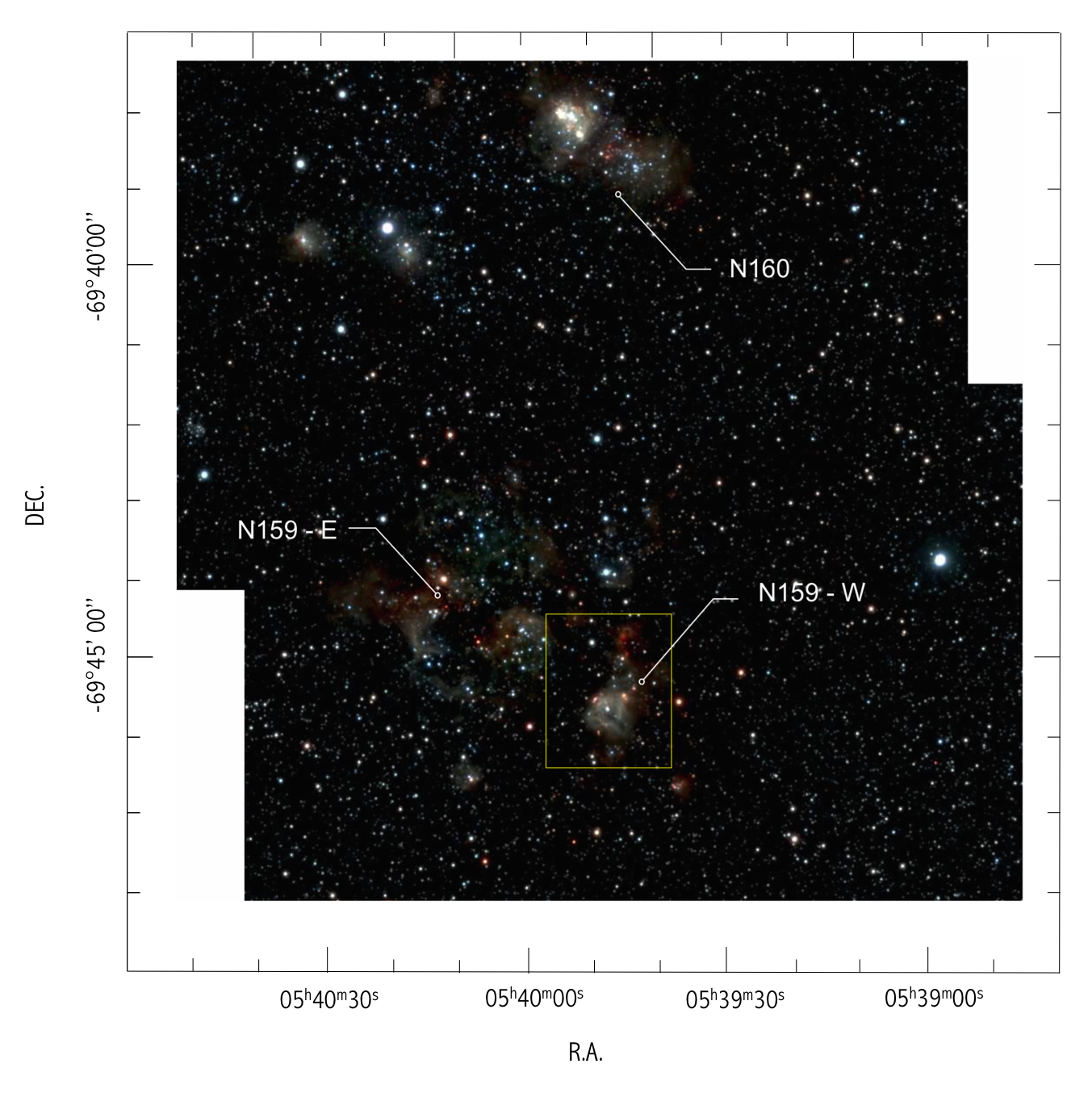}
  \includegraphics[width=8.5cm]{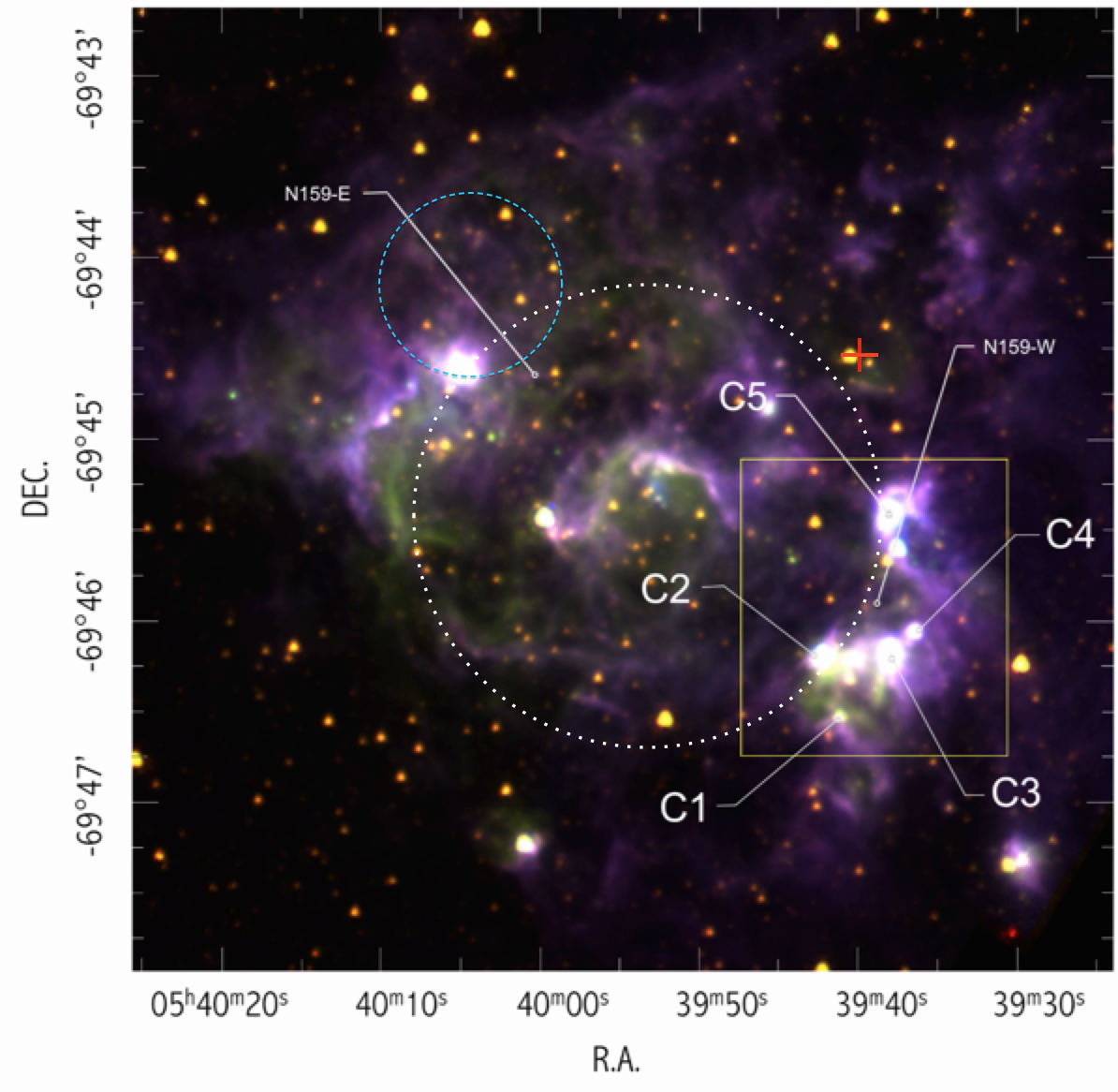}
  \caption[]{Overview of the N159/N160 complex. Top: Composite VISTA
    image \citep{Cioni2011}, combining $Y$ (blue), $J$ (green) and
    \ks\/ (red) bands.  Bottom: Composite {\em Spitzer} image of N159E
    and N159W, combining 3.6\,$\mu$m (blue), 4.5\,$\mu$m (green) and
    8\,$\mu$m (red). The N159W field targetted by GeMS/GSAOI is shown
    by the yellow square. Compact clusters discussed in the paper are
    labelled C1-C5. The positions of the close X-ray binary LMC X-1 and  the SNR 0540 – 697 (\cite{Chen2010a}) are marked in dash-dotted cyan line and red cross respectively.}  \label{fig:vista}
\end{figure}
 
One of the richest star-forming regions in the LMC is the N159/N160
complex (see Fig.~\ref{fig:vista}), with its \hii regions first
catalogued by \citet{Henize1956}. The complex is located in a ridge of
molecular CO gas \citep{Ott2008}, approximately 600\,pc (in
projection) south from the spectacular 30~Doradus region, and is the
location of the first extragalactic protostar, discovered by
\cite{Gatley1981}, as well as the first known type~I extragalactic OH
maser \citep{Caswell1981}. An overview of the structural components of
the complex was given by \citet{Bolatto2000}, who described it as
three distinct, well-separated regions (spanning
$\sim$15$'$/220\,pc), as follows:
\begin{enumerate}
\item {{\it N160:} associated with \hii regions and young stellar
    clusters, where massive-star formation is in a relatively evolved
    stage and the parent clouds are almost entirely photodissociated
    and dissipated.}
\item {{\it N159:} associated with \hii regions and young stellar
    clusters, but still closely linked with molecular gas and active
    star formation. N159 hosts massive embedded young stellar objects
    (YSOs), a maser source, and several ultracompact \hii
    (UCHII) regions \citep{Chen2010a}. N159 harbours
    two giant molecular clouds: N159E and N159W, located on the
    eastern and south-western sides of N159, respectively
    \citep{Johansson1998,Jones2005,Fukui2008}. CO observations also
    reveal a very high concentration of molecular gas at the location
    of N159W \citep{Cohen1988,Johansson1998,Bolatto2000}.}
\item {{\it N159S:} a giant molecular cloud $\sim$60\,pc south of N159
    (thus not included in Fig.~\ref{fig:vista}), with seemingly little
    star-formation activity \citep{Bolatto2000,Galametz2013}, although
    the detection of candidate Herbig Ae/Be stars at its northern tip
    suggests cluster formation might just be starting
    \citep{Nakajima2005}.}
\end{enumerate}

The morphology described above suggests sequential star-formation
southwards through the complex. This is supported by NIR imaging from
\citet{Nakajima2005}, who suggested that star formation in N160 may
have been initiated by a supergiant shell (SGS~19, with its centre
north-northeast of N160) with subsequent star formation in N159 and
the tip of N159S. This scenario is also supported by \cite{Farina2009}
from spectroscopic observations of massive stars across the region.

Early studies of the stellar content and exciting stars from optical
observations of N159 were presented by \citet{HMT1982} and
\citet{Deharveng1992}. From observations with the {\em Spitzer Space
  Telescope} Infrared Array Camera (IRAC), \citet{Jones2005} argued
that the different components in N159 have a common star-forming
history, with a wind-blown bubble ($\sim$1-2\,Myr old) initiating star
formation at the rim. The Papillon Nebula \citep{Meynadier2004} in the
the northeast of the bubble and the C2 and C5 compact clusters (see
below) are located on the edge of this bubble (indicated by the
$\sim$40\,pc diameter circle in the lower panel of
Fig.~\ref{fig:vista}). These arguments are substantiated by the map of
star-formation rate from \citet{Galametz2013}.

From analysis of {\em Spitzer} observations with the Multiband
Infrared Photometer, \cite{Chen2010a} proposed a more complicated
pattern of star formation in N159, where the expansion of the \hii
region could have triggered the star formation in N159E but not in
N159W. They concluded this from a comparison of the distribution of
YSOs and massive stars in both regions, and suggested that star
formation in N159W may have started spontaneously (or was triggered by
a force that is no longer detectable, e.g. an old supernova remnant).

A first high-resolution NIR study of N159W, taking advantage of
adaptive optics (AO), was presented by \citet{Testor2006}, who
resolved some of the candidate YSOs/clusters into multiple components
for the first time. More recently, $^{13}$CO\,(J=2--1) observations of
N159W with the Atacama Large Millimetre Array (ALMA) have discovered
the first two extragalactic protostellar molecular outflows
\citep{Fukui2015}.


In this paper we investigate the stellar content of N159W, using deep,
high angular-resolution {\mbox{$JHK_{\rm s}$}}-band images obtained
with the Gemini South Adaptive Optics Imager
\citep[GSAOI,][]{McGregor2004,Carrasco2011a}, fed by the Gemini
Multi-conjugate adaptive optics System \citep[GeMS,][]
{Rigaut2014,Neichel2014a}.  The observed field covered a field of
$\sim$90$''$ across (as shown in Fig.~\ref{fig:vista}).  These
observations cover a wider field than the previous AO-corrected
imaging, at finer angular resolution, and reach some 2-3\,mag deeper.
This combination of high angular-resolution and excellent sensitivity
in the NIR provides us with an unprecedented view of N159W, to
investigate its stellar content and ongoing star formation.


The paper is organised as follows. In Sect.~2 we present the
observations, data reduction, and completeness and contamination
analysis. Sect.~3 discusses the morphology of the region, and
colour-colour and colour-magnitude diagrams from the data are used to
identify candidate YSOs. In Sect.~4 we derive the luminosity function,
and associated IMF of the central cluster.  Sect.~5 discusses the
star-formation scenario in the N159W complex, with a closing summary
in Sect.~6.


\section{Observations and data reduction}


\subsection{Observations}

\begin{figure*}
   \centering
 \includegraphics[width=\hsize]{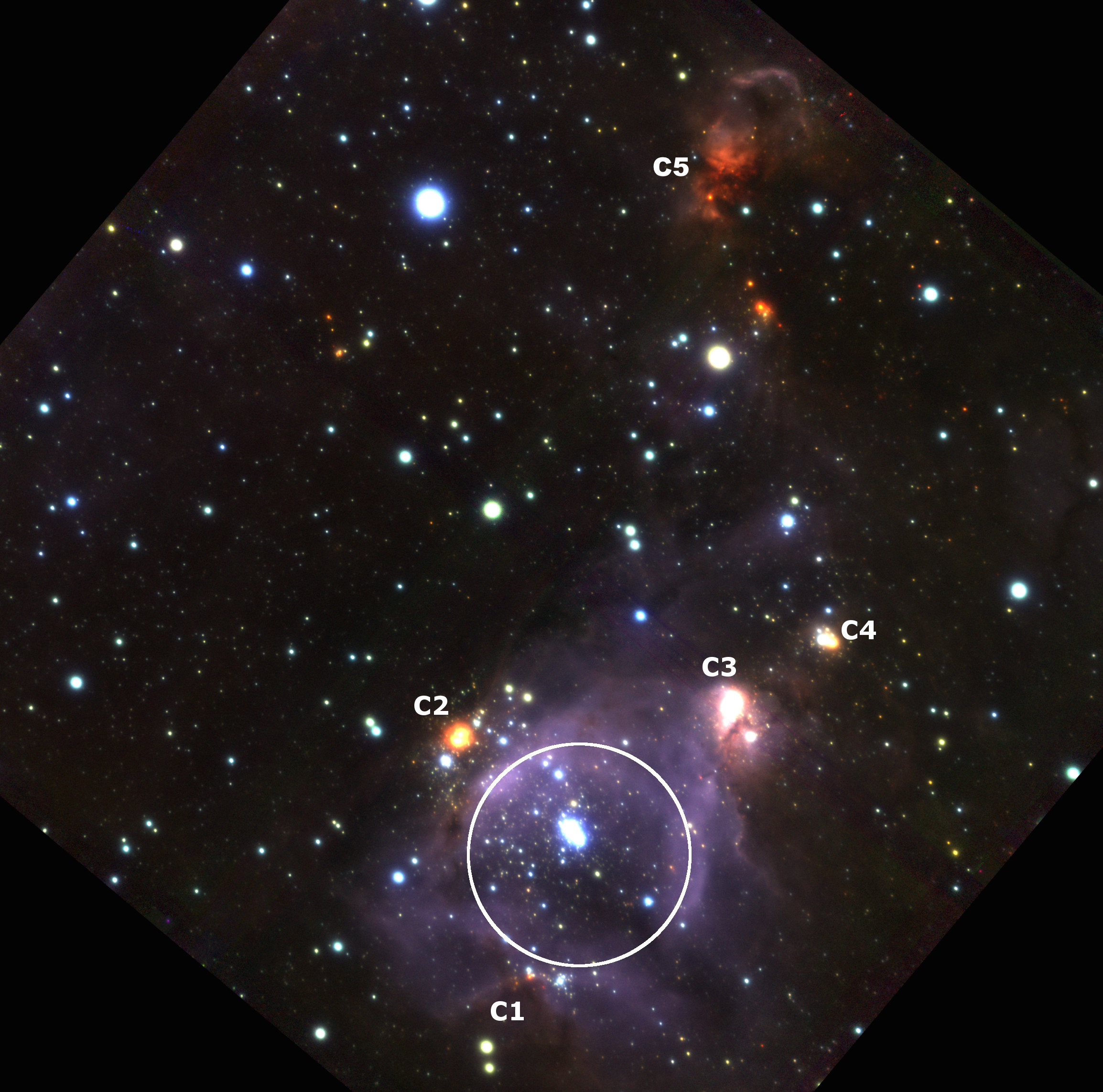}
 \caption{Three-colour GeMS/GSAOI image of N159W combining $J$ (blue),
   $H$ (green), and \ks\/ (red); north is up and east is left. The
   width of the image (east to west) is 90$''$ ($\sim$22\,pc). Compact
   clusters discussed in the paper are labelled C1-C5, and the white
   circle indicates the H\,{\footnotesize II} bubble enclosing the
   central star cluster.}
              \label{fig:threecolor}%
\end{figure*}

The data were obtained with the Gemini South Adaptive Optics Imager
\citep[GSAOI,][]{McGregor2004,Carrasco2011a}, fed by the Gemini
Multi-conjugate adaptive optics System \citep[GeMS,][]
{Rigaut2014,Neichel2014a} on 2014 December 8 as part of
programme GS-2014B-C-2 (P.I. B.~Neichel). GSAOI provides images of a
85$''$\,$\times$\,85$''$ field of view, at a plate scale of
$\sim$20\,mas/pixel \citep[see][]{McGregor2004,Carrasco2011a}.  The
focal plane is imaged by a 2\,$\times$\,2 mosaic of Hawaii-2RG
2048\,$\times$\,2048 pixel arrays, with 3\farcs0 gaps between each
array.

The observations are summarized in Table~\ref{tab1} and the final
three-colour ({\mbox{$JHK_{\rm s}$}}) image is shown in
Figure~\ref{fig:threecolor}. Each science observation was randomly
dithered around the central pointing by 5$''$ to fill-in the
detector gaps, and adjacent sky frames were taken 2\farcm5 from the
science field. The average resolution obtained over the field, as
measured by the full-width half maximum (FWHM) of the stars on
single-exposure frames, is reported in Table~\ref{tab1}, as is the
natural seeing at the time of the observations (measured at zenith).
The average FWHM in the \ks-band was 105\,mas, while the average
Strehl ratio (SR) obtained was 15\,\%. These performances are slightly
worse than those expected for such good-seeing conditions, but are
consistent with observations at low altitude (airmass between 1.3 and
1.6), as is the case for the LMC as seen from Cerro Pachon. The
coordinates of the field centre were $\alpha$\,=\,05$^h$39$^m$40$^s$,
$\delta$\,$=$\,$-$69\degr45\arcmin 55\arcsec (J2000), chosen to span
two star-forming regions of interest and to overlap with the field
observed by \cite{Testor2006}.

\begin{table*}
\begin{center}
  \caption{Observational details and {\sc starfinder} parameters used
    for the star detection and photometry for each filter. The quoted
    full width half maximum (FWHM) and Strehl ratio (SR) of the
    point-spread functions (and their corresponding standard
    deviations) are the means from measurements of over 200 stars
    uniformly distributed over the field, from all the individual
    frames. The exposure time for each image was 80\,s. The two/three
    values given for the {\sc starfinder} threshold mean that the
    algorithm does two/three iterations, with a relative threshold at
    $N$-sigma each.}\label{tab1}
\begin{tabular}{lccccccccc}
\hline
\hline
Date &Filter & Number of & $<$FWHM$>$ & $\sigma_{\rm{FWHM}}$ & $<$SR$>$ & $\sigma_{\rm{SR}}$ & Natural seeing 
& {\sc starfinder} & {\sc starfinder} \\
     &       & frames  &  (mas)  & (mas) & (\%) & (\%) & (@ 0.55$\mu$m) & Threshold & Correlation \\
\hline
2014 Dec 8 & $J$  & 17  & 145  & 15 & $\phantom{1}$3 & 1 & 0\farcs55 & [5,2] & 0.7$\phantom{5}$ \\
           & $H$  & 14  & 100  & 20 & $\phantom{1}$9 & 2 & 0\farcs60 & [5,3] & 0.7$\phantom{5}$ \\
           & \ks  & 17  & 105  & 15 &             14 & 3 & 0\farcs55 & [5, 4, 2] & 0.75 \\                 
\hline
\end{tabular}
\end{center}
\end{table*}


\subsection{Data reduction}
\label{subsec:red}

The data were reduced using the same methods as those from
\citet{Neichel2015a}, using home-made procedures developed in {\sc
  yorick} \citep{Munro1995}. The relevant steps were: (i) creation of a
master flatfield, based on sky-flat images taken during twilight of
the same night; (ii) creation of a master sky-frame based on the
dedicated sky images; (iii) correction of the science frames using the
relevant master flats/skies, as well as for detector non-linearities
and different gains between each detector.

In addition, it was necessary to apply an instrumental distortion
correction. An initial distortion map was derived from GeMS/GSAOI
observations of the Galactic globular cluster NGC\,288, that were
correlated with images from the Advanced Camera for Surveys on the
{\em Hubble Space Telescope (HST)}. This distortion map lead to a
residual positioning accuracy of $\sim$0\farcs2, which is not good
enough for the image quality of the GeMS/GSAOI images here.  Moreover,
this calibration only accounts for the static instrumental distortion
-- a dynamical component, which depends on the constellation of
natural guide stars and environmental factors like the telescope
pointing, still needs to be taken into account.

The dynamical contribution to the distortion map has to be corrected
frame by frame, selecting one frame as the primary astrometric
reference. By using high-order polynomials and the position of
relatively bright stars common to all the frames, we were able to
cross-register the images by the following steps: (i)
creation of a polynomial base $(P)_i$ of distortion modes as
$P_i(x,y)= C_{i,0} + C_{i,1}x + C_{i,2}y + C_{i,3}x^2 + C_{i,4}xy +
C_{i,6}y^2 + ... $~; (ii) measurement of the vectors $V_x$ and $V_y$
of the positional difference between each image and the reference frame,
for each of the bright ($<$\,21\,mag) stars common to all frames;
(iii) perform a Levenberg--Marquardt fit of the $V_x$ and $V_y$
vectors on the polynomial base and determination of the $\alpha_i$ and
$\beta_i$ coefficients such that: $V_x=\sum \alpha_i P_i $ and $
V_y=\sum \beta_i P_i$; (iv) application of the inverse distortion map
to the full image. 

Of course, the quality of the above corrections depends on the number of
stars available for the procedure. The N159W observations provided a large
number of reference stars and, by using 15 degrees of freedom per axis, we
obtained corrected frames with a typical precision of 0.1\,pixel. Following
the procedure described above, each individual image was reduced
and combined by filter to produce three final reduced images.


\subsection{{\sc starfinder} photometry}
\label{subsec:starfinder}

Photometry for the stars in the observed field was obtained using the
{\sc starfinder} package \citep{Diolaiti2000}. Because of the complex
structure of AO-corrected point spread functions (PSFs), and their
spatial and temporal variations, photometry of AO images requires
PSF-fitting algorithms that provide more accurate results than regular
aperture photometry \citep[see, e.g.][]{Neichel2015a}.

{\sc starfinder} builds an empirical PSF by combining results for
several stars in an image. In the case of spatial variations of the
PSF across a field, it is usually divided into subfields, with {\sc
  starfinder} then used on each subfield in turn. There is naturally a
trade-off between having enough stars in each subfield so that an
accurate PSF model can be extracted, versus keeping the subfield as
small as possible to minimize PSF variations.  We experimented with a
range of subfield sizes for N159W.  The minimum residual was obtained
for subfields of 16$''$\,$\times$\,16$''$, shifted by 8$''$ (in x or
y) to introduce an overlap between the frames. Each star was then
measured several times, using different sets of PSFs, which were then
used to investigate the photometric errors.

The other critical aspect of tuning the {\sc starfinder} analysis is
the choice of the number of iterations, the relative threshold, and
the correlation threshold for star extraction at each iteration. These
parameters are summarized in Table~\ref{tab1}. Note that {\sc
  starfinder} goes through several iterations of star detection,
removing the detected stars at each iteration. We can then adjust
different threshold levels, pushing toward low signal-to-noise over
the iterations. The numbers given in the ninth column of
Table~\ref{tab1} are the thresholds used for each iteration.

We adjusted the {\sc starfinder} parameters so that single faint stars
were detected without too much contamination from remaining bad/hot
pixels or from structures in the background. Our three final reduced
images ($J$, $H$, and \ks) have different characteristics in terms of
noise and resolution, so the {\sc starfinder} parameters have been
optimized for each filter. To detect as many of the faint stars as
possible we used 2\,$\times$\,2 binned images. This enabled the
detection of faint stars more robustly and, as the pixel size is
20\,mas and the typical resolution was $\sim$100\,mas, the binned
images are still well sampled, with no loss of spatial information.
Three bright stars were saturated in the $H$- and \ks-band
observations (with pixel counts in excess of 50\,000\,e$^-$). These
were excluded from the PSF definition and photometry, with their
magnitudes adopted from observations of N159/N160 with the Infra-Red
Survey Facility (IRSF) by \citet{Nakajima2005}. The final
photometric catalogue contains magnitudes for 2185 stars in $J$, 4187
in $H$ and 3912 in \ks.

\subsection{Photometric and astrometric calibration}
\label{subsec:zp}

To calibrate the instrumental magnitudes we evaluated the photometric
zero-points, employing values for 40 well-isolated stars in the
observed field from the IRSF observations \citep{Nakajima2005}.  A
reasonable match is found, with a zero-point uncertainty of 0.06\,mag
in $J$ and $H$, and 0.08\,mag in \ks.

As an independent check, we also attempted to derive a zero-point
using the 2MASS catalogue \citep{Skrutskie2006}. The pixel size of the
2MASS cameras was only $\sim$2$''$, so there is a large mismatch
compared with the GSAOI images, and only 10 of the isolated stars used
above could be securely identified. The zero-point uncertainties from
calibration with the 2MASS stars were 0.2\,mag in $J$ and $H$, and
0.1\,mag in \ks.  Given the smaller number of stars availale, we
therefore adopted the zero-points determined using the IRSF
catalogue\footnote{Corrections between the IRSF and 2MASS systems are
  small, amounting to $<$\,0.05\,mag for ($J$\,$-$\,\ks)$_{\rm
    IRSF}$\,$<$\,1.7\,mag \citep{Kuc2008}, so are neglected when
  estimating the final photometric error.}.

The world-coordinate system was also calibrated using the same 40
isolated stars from the IRSF catalogue \citep[for which the quoted
astrometric accuracy is 0.5$''$;][] {Nakajima2005}. The astrometric
solution of the GSAOI images was set using the x/y (pixel) positions
of each star cf. the IRSF positions, resulting in a mean astrometric
dispersion of 5\,mas (i.e. less than the typical FWHM of the images,
so no confusion is expected).


\subsection{Photometric error and completeness}
\label{subsec:comp}

After calibration, we undertook a series of completeness tests in each
filter to estimate the depth and photometric uncertainties of our
images. The detectability of a source depends on the local stellar
density and the {\sc starfinder} settings, and may also be affected by
diffuse background in the case of a region like N159W. To determine
the completeness of our images we used simulated stars inserted in the
real images. As for the PSF analysis, we divided the image in
subfields and used the corresponding local PSF from {\sc starfinder}
to construct simulated stars which mimic the real variations across
the field.

A set of 1000 artificial stars was inserted at random positions in
each subfield, following the method from \citep{Neichel2015a}. Random
magnitudes (in steps of 0.5\,mag) between 14.0 and 22.0 were assigned 
for each filter. The {\sc starfinder} source detections were then run
using the exact parameters as those in Table~\ref{tab1}, and the number
of artificial stars successfully recovered was used to estimate the
photometric completeness. At the same time we also estimated
the photometric error by computing the standard deviation of the
difference between the measured and simulated magnitudes. 

\begin{figure}
\begin{center}
\begin{minipage}[c]{.45\textwidth}
\includegraphics[width=\hsize]{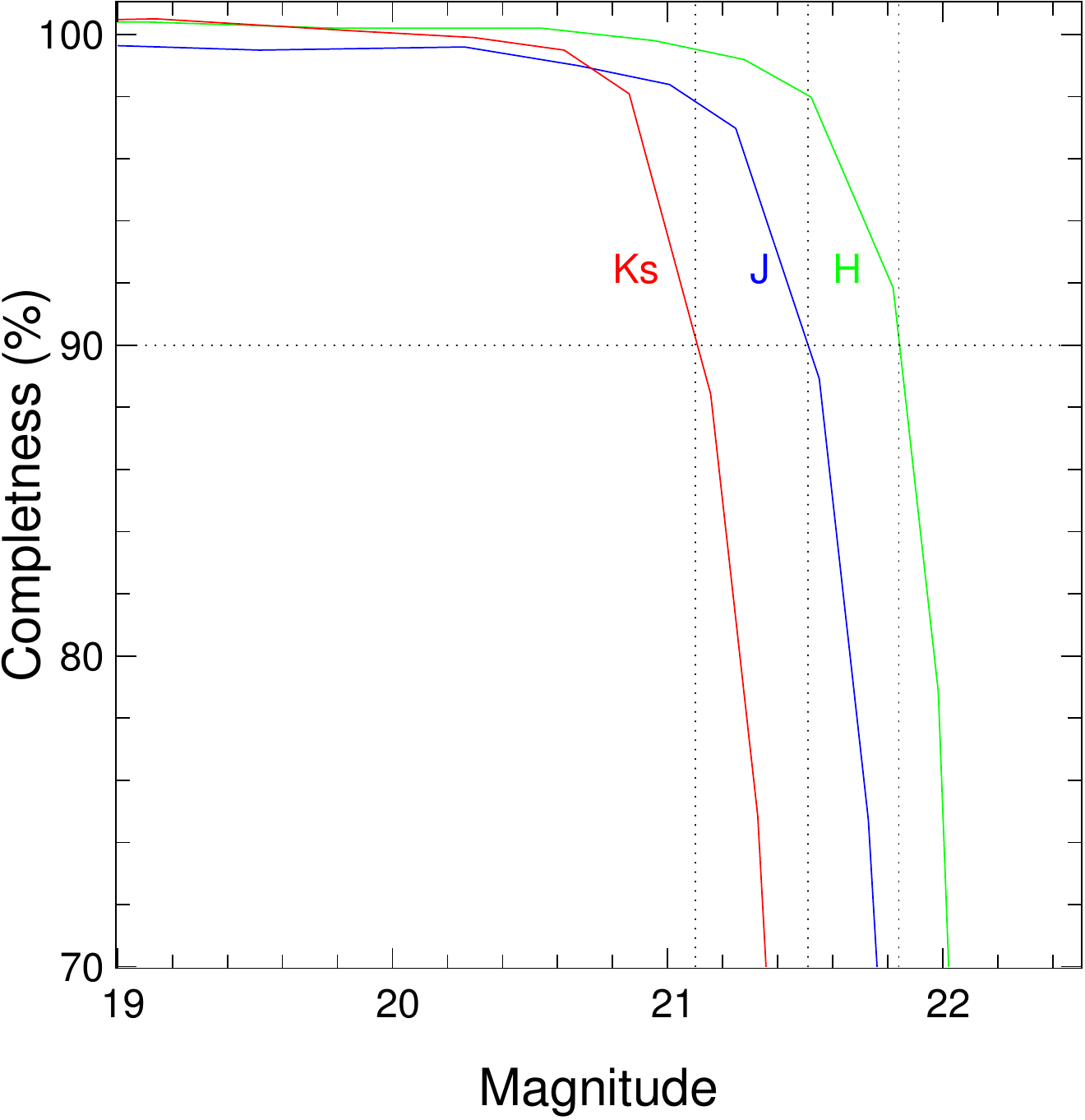}
\includegraphics[width=\hsize]{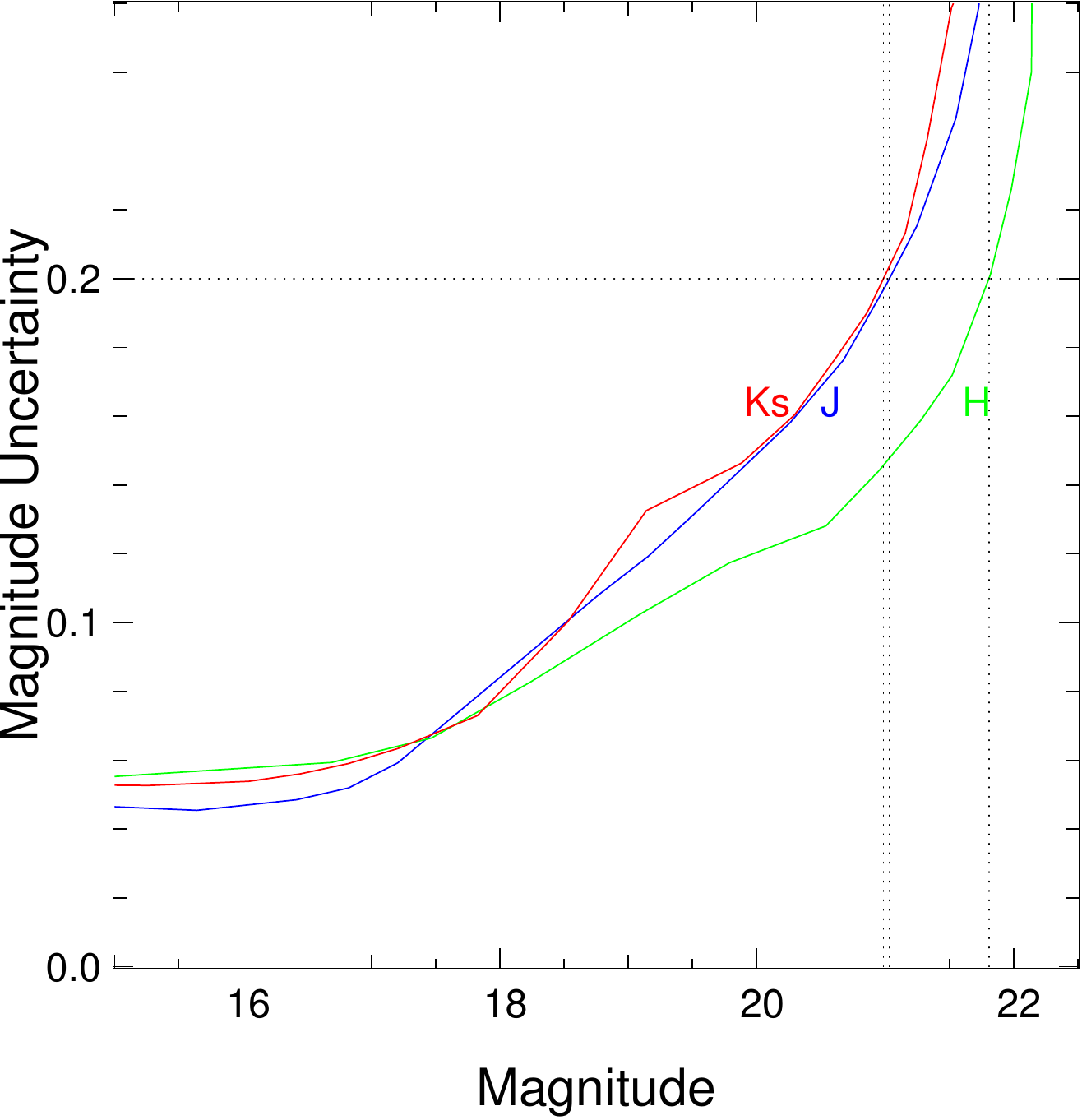}
\end{minipage}
\caption{Completeness (upper panel) and photometric uncertainties (lower panel) for
  each filter estimated from the completeness tests.}\label{fig:mag_uncert}
\end{center}
\end{figure}

Results of these tests are shown in the upper panel of
Fig.~\ref{fig:mag_uncert}, with 90\,\% completeness obtained at
$J$\,$=$\,21.55, $H$\,$=$\,21.85, and
\ks\,$=$\,21.85\,mag\footnote{This is 2.55, 3.85, and 4.50\,mag deeper
  than the respective limiting \jhk magnitudes of the IRSF data
  \citep{Nakajima2005}.}. Results for the photometric uncertainties
are shown in the lower panel of Fig.~\ref{fig:mag_uncert}; the final
photometric uncertainties are given by these errors combined in
quadrature with the zero-point errors estimated above.

\section{Results}

\subsection{Morphology of the N159W complex}
\label{ssec:morphology}

The composite GeMS/GSAOI image of N159W is shown in
Fig.~\ref{fig:threecolor}. Toward the south of the observed field we
can clearly identify the ionised \hii region, forming an almost
circular nebula (as indicated by the white circle in the figure).
There are two bright stars separated by $\sim$0\farcs9 at the centre
of the nebula, with $J$\,$=$\,14.15 and 14.85\,mag (with the northern
star the brightest, as shown in the lower-right panel of Fig.
\ref{fig:objA7}); these are thought to be the primary ionising sources
of the \hii region \citep{Deharveng1992}.  Optical spectroscopy from
\cite{Conti1991} classified the (then unresolved) composite spectrum
as O3-6~V, while also suggesting that it might well be a small compact
cluster. \cite{Deharveng1992} classified the two stars as O5-6 and
O7-O8, and \cite{Testor2006} obtained a NIR spectrum of the brightest
star, confirming it as an O star of type O5-6, with an estimated
extinction of \av = 3.54\,mag.  For completeness, we note that
\cite{Farina2009} classified their spectrum of the central source(s)
as O8\footnote{Their observed coordinates are $\sim$0\farcs3 from the
  position of the brighter star so their spectrum was likely dominated
  by this object, albeit contaminated by the other star given their
  quoted pixel scale and slitwidth.}. The cluster associated with
these two massive stars, within the surrounding nebula, is hereafter
referred to as the central cluster.

Three compact clusters are observed at the edge of the nebula,
identified as C1, C2 and C3 in Fig.~\ref{fig:threecolor}. These are
located symmetrically around the nebula, at an almost equidistant
projected distance from the centre. Another compact cluster
(identified as C4) is located to the north-east of C3, and further
north there is a very embedded region, labelled as C5. The properties
of each of these are now discussed in turn.

\subsubsection{Compact cluster C1}
This corresponds to a class~II YSO (053940.78$-$694632.0) identified
by \citet{Chen2010a}, which was also classified as a Herbig Ae/Be star
by \citet{Nakajima2005}. Our observations reveal a compact cluster
(see Fig.~\ref{fig:objA7}), with the reddest star consistent with the
position of the previously reported YSO/Herbig object. The clump that
hosts this cluster has a somewhat cometary shape, with a
limb-brightened rim, and is further discussed in
Sect.~\ref{sec:scenario}.

\subsubsection{Compact cluster C2}
This compact cluster is source P2 from \citet{Jones2005}, and N159A7
from \citet[][resolved by their observations into their sources 121,
121a, 123, and 123a]{Testor2006}. This also corresponds to a
class~I/II YSO (053941.89$-$694612.0) from \cite{Chen2010a}, who
estimated its mass to be 33.7\,$\pm$\,2.6\,\msun, with
\av\,$=$\,12.7\,mag (while also noting it had multiple components).

\begin{figure*}
   \centering
 \includegraphics[width=\hsize]{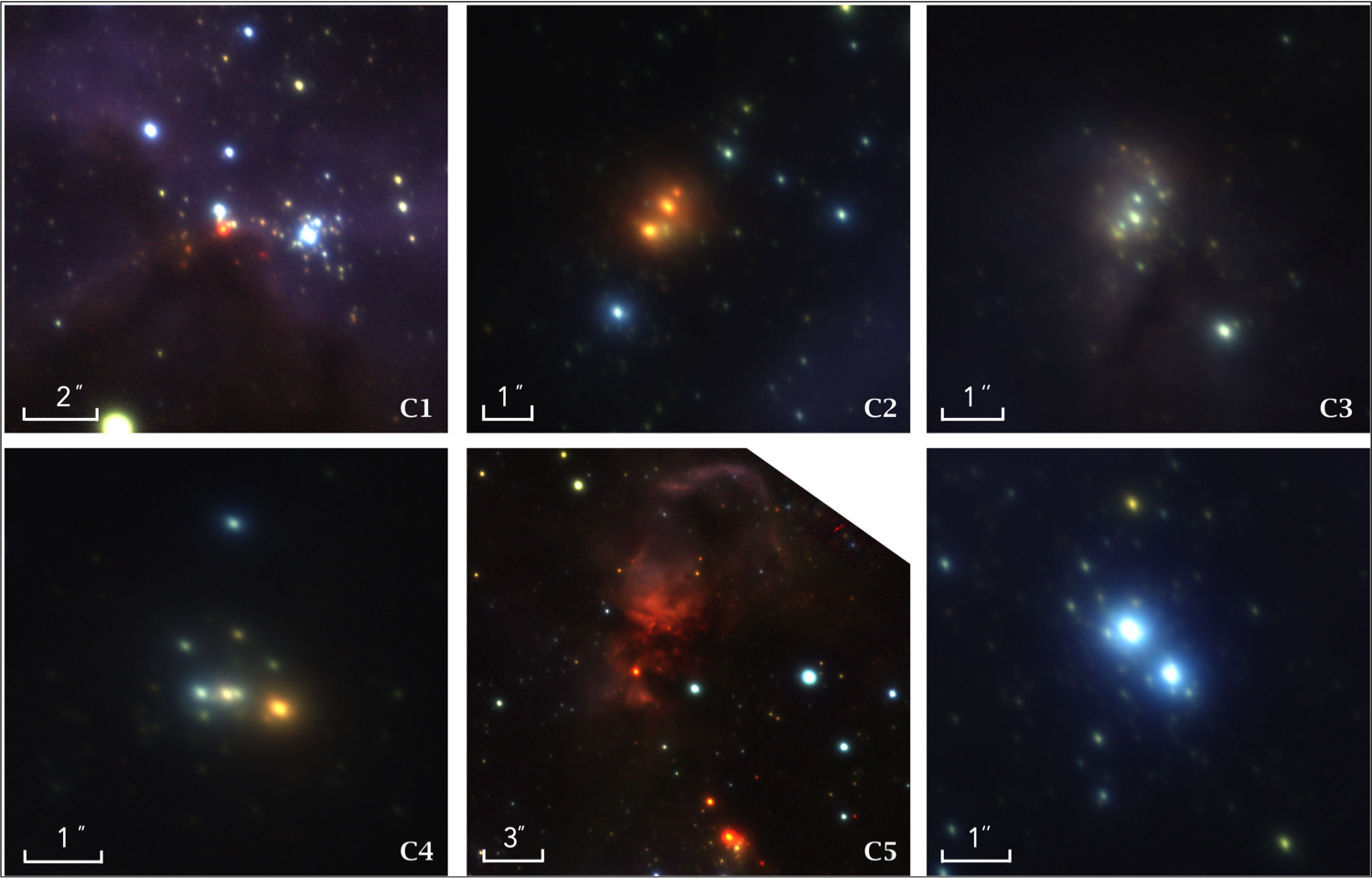}
 \caption{Details of the five compact clusters (C1-5) and central two
   O-type stars (thought to ionise the central cluster) from the
   three-colour GeMS/GSAOI image; $J$ (blue), $H$ (green),
   and \ks\/ (red).}\label{fig:objA7}
\end{figure*}

The expanded view of C2 from the GeMS/GSAOI images is shown in
Fig.~\ref{fig:objA7}, revealing two dominating components and several
other sources, all embedded in a diffuse nebula with a diameter of
$\sim$0.55\,pc. The two main components have comparable \ks-band
magnitudes, while the redder source (123 from \citeauthor{Testor2006})
is probably the source dominating the YSO luminosity and mass estimate
from \citeauthor{Chen2010a}; spectroscopy of this source from
\citeauthor{Testor2006} confirms that it is in a very early stage
(class~0/I YSO), so it is unlikely to have started ionising its
surroundings. Interestingly, C2 is one of the two high-mass sources
with a reported outflow from the ALMA observations by
\citet{Fukui2015}. They suggest that star formation in this compact
cluster was likely triggered by the collision of two thin filaments of
gas $\sim$10$^5$\,yrs ago.

\subsubsection{Compact cluster C3}
This cluster was detected as a compact radio source by \citet[][their
source~5]{Hunt1994} and \cite{Indebetouw2004}. It corresponds to
053937.53$-$694609.8 from \citet{Chen2010a}, who classified it as a
type I/II YSO (with multiple components). Their mass estimate was
31.2\,$\pm$\,2.9\,\msun, with an inferred spectral type of O7~V (cf.
the spectral type of O7.5~V inferred from the analysis of radio
data by \citeauthor{Indebetouw2004}). C3 is also N159A5 from
\citet{Testor2006}, who described it as a tight cluster of 
five stars (their nos. 133, 136, 137, 138 140), with an estimated
extinction of $A_V$\,$=$\,8-12\,mag.

All the components are embedded in nebulosity with a diameter of
1\farcs9 (0.5\,pc). Our GeMS/GSAOI data resolve C3 into even more
components than past observations with $\sim$20 sources (see
Fig.~\ref{fig:objA7}). \citet{Jones2005} described the cluster as a
compact \hii region with one or more OB-type stars, and spectroscopy
from \citet{Testor2006} of the brightest member (hence likely dominant
source of ionising photons) revealed a late O-type star \citep[also
see][]{Martin-Hernandez2005}.

\subsubsection{Compact cluster C4}
This corresponds to N159A6 from \citet{Testor2006}, who resolved it
into six stars embedded in nebulosity with a diameter of 0.45\,pc.
Our observations resolve C4 into at least 12 stars, as shown in the
lower-left panel of Fig.~\ref{fig:objA7}. In particular, star 155 from
\citeauthor{Testor2006} (in the centre of the image) is now resolved
into a double star. C4 is also source 053935.99$-$694604.1 from
\citet{Chen2010a}, who classified it as a type I YSO and noted it
having multiple components. Indeed, using the photometry from
\citeauthor{Testor2006} (their star 151, the bright, red object on the
western edge of C4 in Fig.~\ref{fig:objA7}), \citeauthor{Chen2010a}
estimated a mass of 18.5\,$\pm$\,1.4\,\msun, with
$A_V$\,$=$\,2.5\,mag; this object was also classified as a Herbig
Ae/Be star by \cite{Nakajima2005}.

\subsubsection{Compact cluster C5 }

C5 is a very red region which appears to be deeply obscured with
nebulosity detected fin the $H$- and \ks-bands (see
Figs.~\ref{fig:threecolor} and \ref{fig:objA7}). It is the most
intense 6\,cm (4.8\,GHz) continuum source seen in observations with
the Australia Telescope Compact Array (ATCA) of N159
\citep{Hunt1994,Indebetouw2004}, and has the highest dust
surface-densities in the whole N159/N160 complex, and one of the
highest star-formation rates \citep{Galametz2013}.

The {\em Spitzer} observations from \citet{Chen2010a} revealed the
presence of an embedded YSO in C5, which they classified as type~I,
estimating a mass of 34.8\,$\pm$\,8.4\,\msun, with an inferred
spectral type of O6~V. More recently, ALMA observations of the dense
CO gas in C5 revealed a complex structure \citep[their
`N159W-N';][]{Fukui2015}, with several filaments which are generally
elongated in the northeast-southwest direction, and the detection of a
blue-shifted lobe of an outflow, thought to be associated with the YSO.
Taking into account longer-wavelength data from the {\em Herschel Space
  Observatory}, \citeauthor{Fukui2015} estimated the mass and
luminosity of the YSO (their `YSO-N') to be 31\,$\pm$\,8\,\msun\/ and
(1.4\,$\pm$\,0.4)\,$\times$\,10$^5$\,\lsun, respectively.

No stars are detected in our GeMS/GSAOI images at the location of the
YSO \citep{Chen2010a,Fukui2015} or the UCHII region
\citep{Indebetouw2004}. Indeed, only one very red star (with
$J$\,$=$\,21.9, $H$\,$=$\,18.4 and \ks\,$=$\,16.2\,mag) is detected
in/near the clump, some 2$''$ (0.5\,pc) south of the UCHII region.

\subsection{Identification and distribution of young stellar sources}

\subsubsection{Near-IR colour-colour diagram}
\label{subsec:ccd}

\begin{figure*}
\centering{
\includegraphics[width=8.5cm,height=8.5cm]{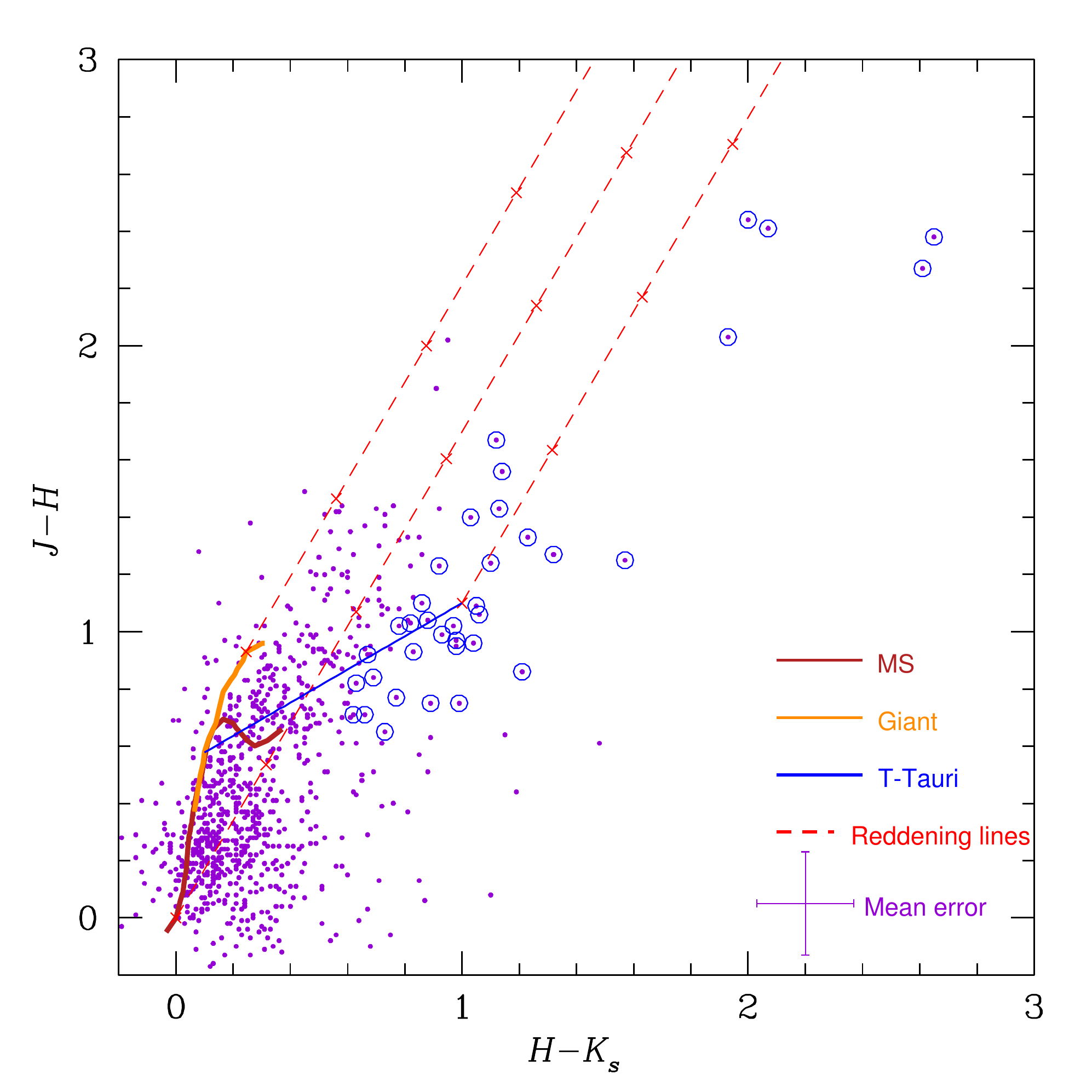}
\includegraphics[width=8.5cm,height=8.5cm]{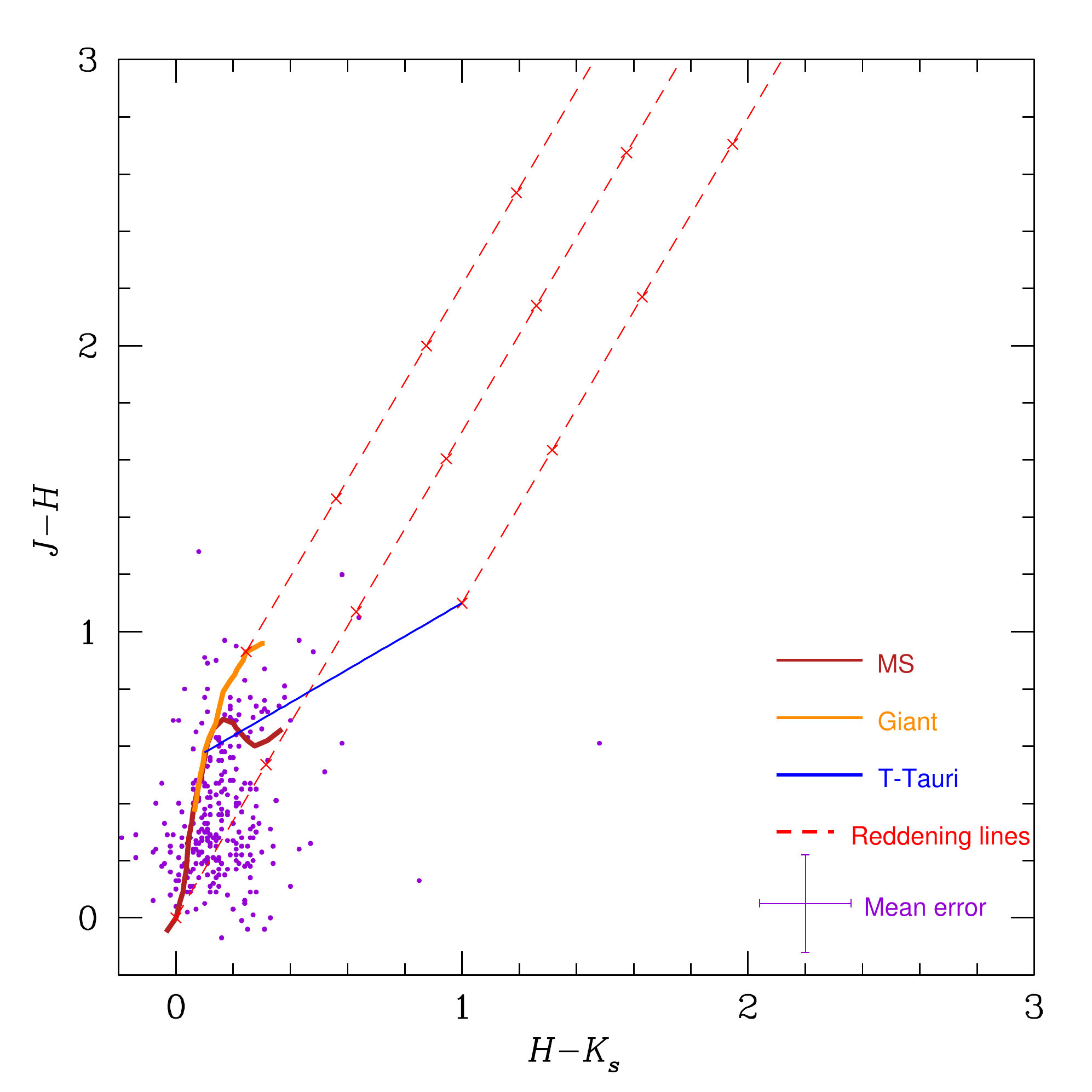}}
\caption{Colour-colour diagrams for the cluster region in N159W
  (left-hand panel) and the control region (right), with the loci of
  the main sequence (MS), giant branch, and T~Tauri stars (TTS)
  overlaid in red, orange and blue, respectively. Also shown are
  reddening vectors (dashed red lines) drawn from the tip of the giant
  brach (left), from the base of the MS (middle), and from the reddest
  TTS (right). Cluster sources with a NIR excess are indicated in the
  left-hand panel by the blue circles. The error bars in the
  lower-right corner show the average uncertainties in the colour
  terms.}
\label{fig:cc-diag}
\end{figure*}

To investigate the nature of the stellar population of N159W, in the
left-hand panel of Fig.~\ref{fig:cc-diag} we show the $(H-\ks)$ vs.
$(J-H)$ colour-colour diagram (CCD) for the 1386 sources detected in
each of the $JHK_{\rm s}$ bands (for which the photometric error is
lower than 0.2\,mag). Overlaid on the CCD are curves
representing the main sequence (MS) and giant branch \citep[shown in
red and orange, respectively;][]{Bessell1988}, and the locus of
T~Tauri stars \citep[TTS, shown in blue;][]{Meyer1997}\footnote{In
  which the MS and TTS loci have been converted to the 2MASS system
  using the relations from \cite{Carpenter2001}.}. 

Reddening vectors are also overplotted in Fig.~\ref{fig:cc-diag} (as
dashed red lines), drawn from the tip of the giant branch (left), from
the base of the MS (middle), and from the reddest TTS (right). In each of
these we calculate the reddening adopting a normal extinction law,
with: $A_J/A_V$\,$=$\,0.282, $A_H/A_V$\,$=$\,0.175,
$A_K/A_V$\,$=$\,0.112 \citep{Rieke1985}. Shifting the brightest stars
along the reddening vector to the MS, we find that they are consistent
with the expected colours of O4-O5 stars, in good agreement with the
spectroscopy of the central stars discussed above.

Most of the sources in the left-hand panel of Fig.~\ref{fig:cc-diag}
are within the reddening band of the MS, which could be a mixture of
reddened field stars and sources with warm circumstellar dust, a
standard characteristic of young pre-main-sequence (PMS) objects
\citep{Lada1992}. In contrast, stars lying redwards of the middle
reddening vector are expected to be associated with NIR-excess from
optically-thick circumstellar disks. We can therefore use the CCD
for a first identification of NIR-excess objects.

To understand the potential contamination from background/foreground
stars, the same CCD is shown for the control field in the
right-hand panel of Fig.~\ref{fig:cc-diag}. Most of the sources are
also within the MS reddening band, and only a few objects are found further
redwards, which could be due to unresolved binaries.

Thus, to try to identify candidate YSOs which are associated with the
cluster we applied the following two criteria:
\begin{enumerate}
\item{$H-\ks$\,$>$\,1\,$\sigma$ (where $\sigma$ is
the error on the colour term) from the MS reddening vector.}
\item{$J-H$\,$>$\,0.65\,mag to avoid the region in the control field
which contains other sources.}
\end{enumerate}
We may have missed a few candidates using these criteria, but we aimed
to minimise the contamination by non-cluster members in the final
sample. The 36 candidate YSOs with NIR excesses identified using these
criteria are indicated in Fig.~\ref{fig:cc-diag} with blue circles.

We comment that 2256 sources were detected only in the $H$- and \ks-bands;
while these are not included in Fig.~\ref{fig:cc-diag}, some may also
be YSO candidates with NIR excesses. Equally, some YSOs appear
relatively normal in the NIR, but are found to display excesses when
going to even longer wavelengths, e.g. from {\em Spitzer} observations
\citep{Ojha2011,Chavarria2014,Saral2015}. Thus, the number of YSO
candidates identified from our analysis is knowingly an underestimate
of the likely total in the region. Either deeper $J$-band observations
and/or high spatial-resolution $L$- and $M$-band observations are
needed to recover all the candidates.  Nevertheless, in the next
section we attempt to recover some of these embedded sources from
their $H-\ks$ colours.


\subsubsection{Near-IR colour-magnitude diagram}
\label{subsec:cmd}

\begin{figure}
\centering{
\includegraphics[trim=0cm 0cm 0cm 2cm, width=11cm,height=9.8cm]{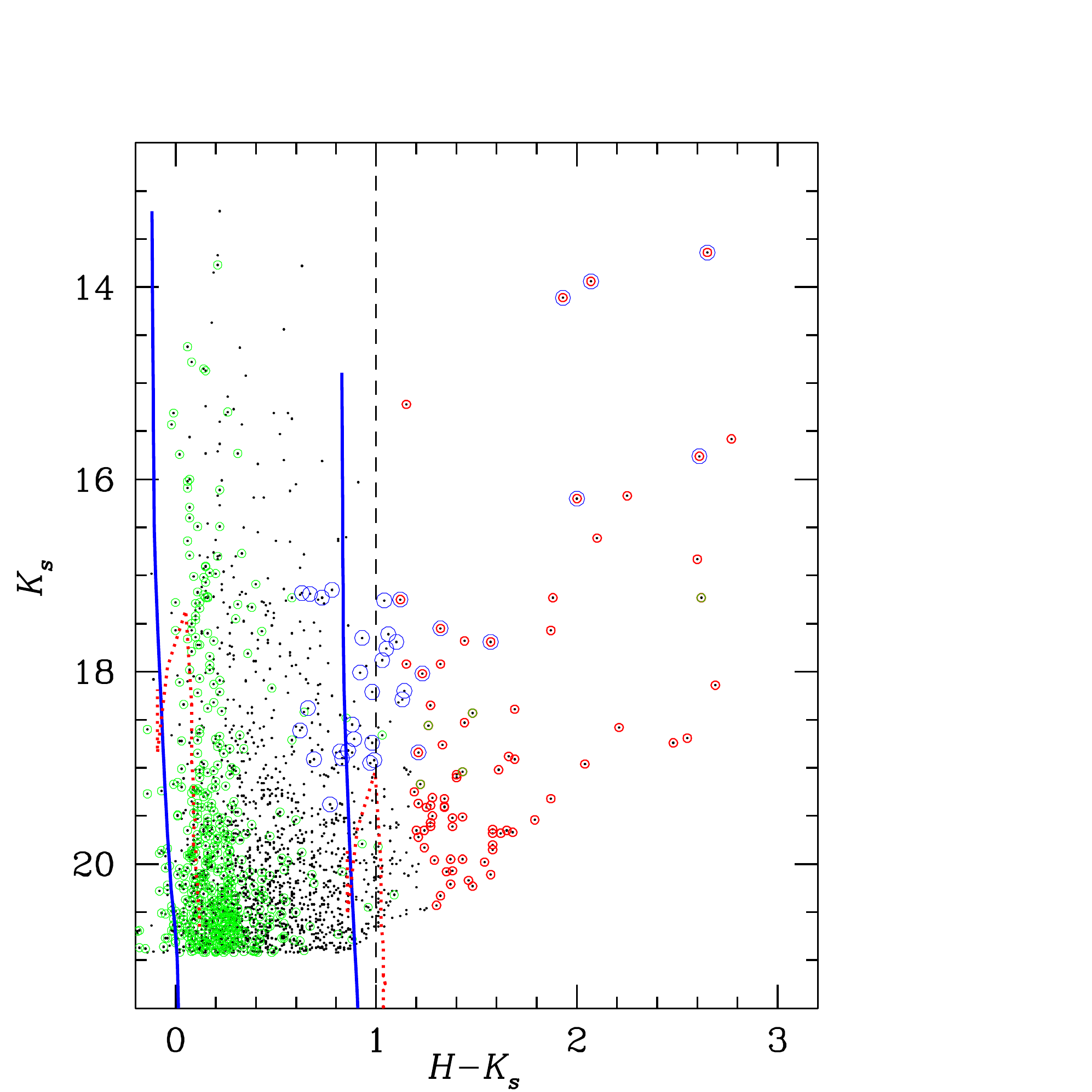}}
\caption{Colour-magnitude diagram of sources in N159W detected in only
  the $H$- and \ks-bands (black points) and those in the control field
  (black points with green circles). Also plotted are the zero-age main sequence (solid
  blue lines) and the 1\,Myr PMS isochrone from \citet[][dotted red
  lines]{Siess2000} for extinctions of A$_V$\,$=$\,0 and 15\,mag.  Red
  circles highlight candidate YSOs identified from our colour cut (see
  text for details). Blue circles highlight candidate YSOs previously identified from the CCD (Fig. \ref{fig:cc-diag}).}\label{fig:KHK}
\end{figure}

The colour-magnitude diagram (CMD) of the 2256 sources detected only
in the $H$- and \ks-bands (with photometric errors of $<$\,0.2\,mag)
is shown in Fig.~\ref{fig:KHK} (black points), together with stars
observed in the control field overlaid (black points with green circles). Also plotted
are the zero-age main sequence (solid blue lines) and the 1\,Myr PMS
isochrone from \citet[][dotted red lines]{Siess2000}\footnote{We
  employed the isochrones from \citet{Siess2000} as they are the only
  PMS evolutionary models available which extend to a mass limit of
  0.1\,\msun\/ and with a subsolar metallicity (calculated for
  $Z$\,$=$\,0.01, cf. the slightly lower metallicity of the LMC of
  $Z$\,$=$\,0.008).}  for extinctions of A$_V$\,$=$\,0 and 15\,mag.
In contrast with the control field, N159W has a significant number of
stars with $(H-K)$\,$>$\,0.7\,mag, which could be associated with the
young population of the cluster. However, trying to disentangle young
stars from reddened background stars from such a CMD is complex, so we
adopted the following approach. We considered the maximal extinction
seen in the direction of N159W, and applied this to the intrinsic
colour of a dwarf star. We then considered all sources redder than
this as potential NIR-excess sources associated with the cluster
region.

We estimated the maximum extinction using the 21\,cm H\,{\footnotesize
  I} ATCA observations from \citet[][which had an angular resolution
of $\sim$7$''$]{Dickey1994}.
From the H\,{\footnotesize I} absorption in the direction of N159W
they estimated N(H\,{\footnotesize
  I})\,$=$\,9.62\,$\times\,$10$^{22}$\,\cmsq. The gas-to-dust ratio,
$N_{\rm H}/E(B-V)$, in the LMC is several times larger than the
Galactic value \citep{Lequeux1989,Clayton1985}. Using the conversion
$N_{\rm H}/E(B-V)$ = 2 $\times $ 10$^{22}$ atoms \cmsq mag$^{-1}$ from
\cite{Lequeux1989} and \rv = $A_V/E(B-V)$ = 3.1, we estimated a
maximal visual extinction toward N159W of
$A_V$\,$\sim$15\,mag\footnote{This agrees with the maximum value from
  the YSO analyses by \citet{Chen2010a}}. The resulting $E(H-K)$ for a background star
is $\sim$0.9\,mag \citep[from
$A_V$\,$=$\,15.9\,$\times$\,$E(H-K)$,][]{Rieke1985}. Considering the
intrinsic $H-\ks$ colour of 1\,Myr PMS stars is $\sim$0.1\,mag (see
Fig.~\ref{fig:KHK}), $A_V$\,$=$\,15\,mag would shift their location to
maximum of $(H-\ks)$\,$\sim$\,1\,mag. 

Thus, we considered sources with $H-\ks$\,$>$\,1\,mag as candidate
YSOs, with intrinsic excesses from disks and/or envelopes.  In our
sample selection we employed a slightly more conservative colour cut,
only selecting those sources with $H-\ks$\,$>$\,1\,$+$\,$\sigma$,
where $\sigma$ was the estimated uncertainty on the colour term.
Applying this method we identified 68 additional candidates, as shown
by red circles in Fig.~\ref{fig:KHK}, although we acknowledge that a
small number of these might be highly-reddened background sources.
This gives 104 candidate YSOs in total: 36 from analysis of the CCD,
and 68 additional candidates from the CMD.


\subsubsection{Spatial distribution of candidate YSOs}
\begin{figure*}
\begin{tabular}{cc}
\includegraphics[width=8.5cm,height=8.5cm]{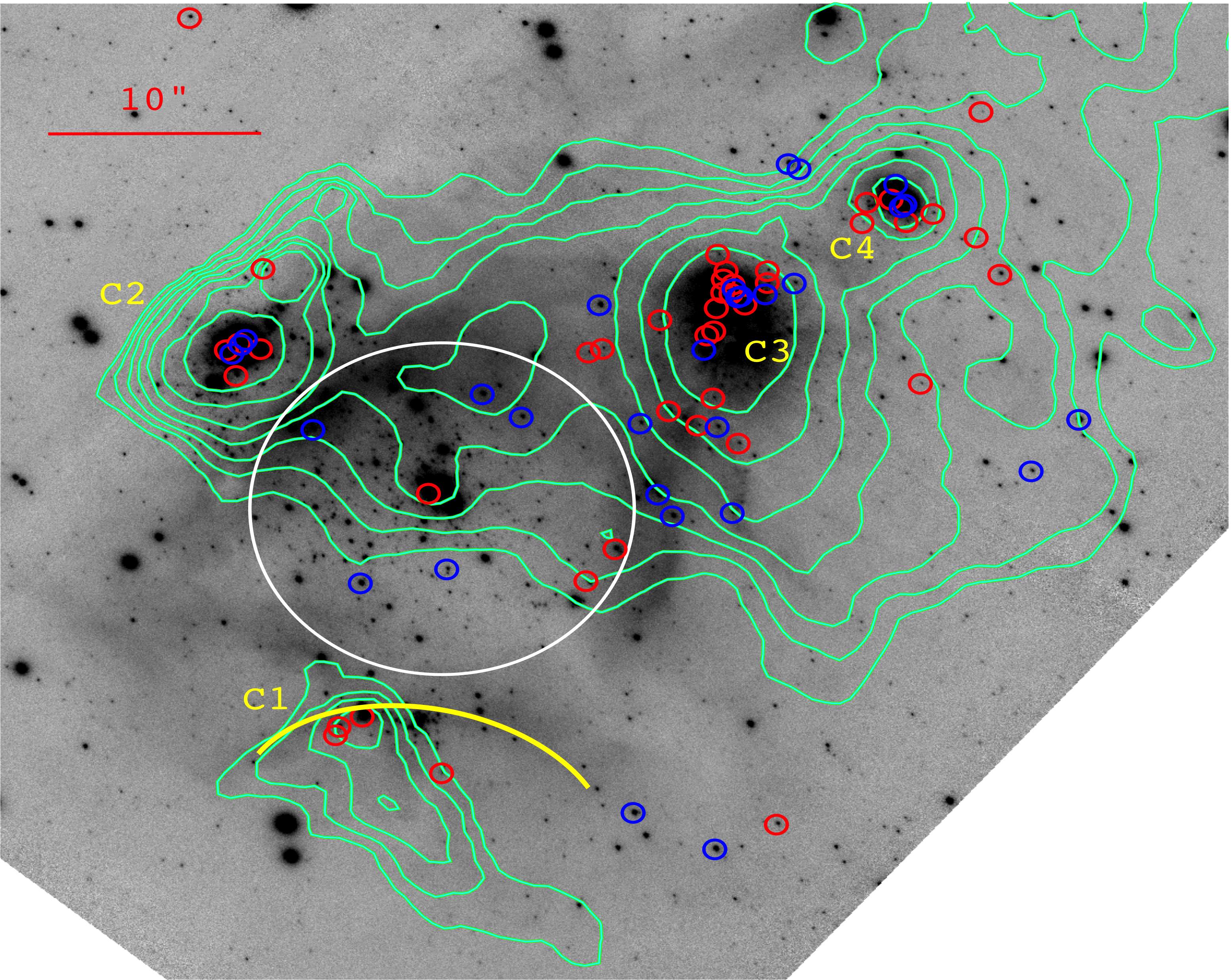} &
\includegraphics[width=8.5cm,height=8.5cm]{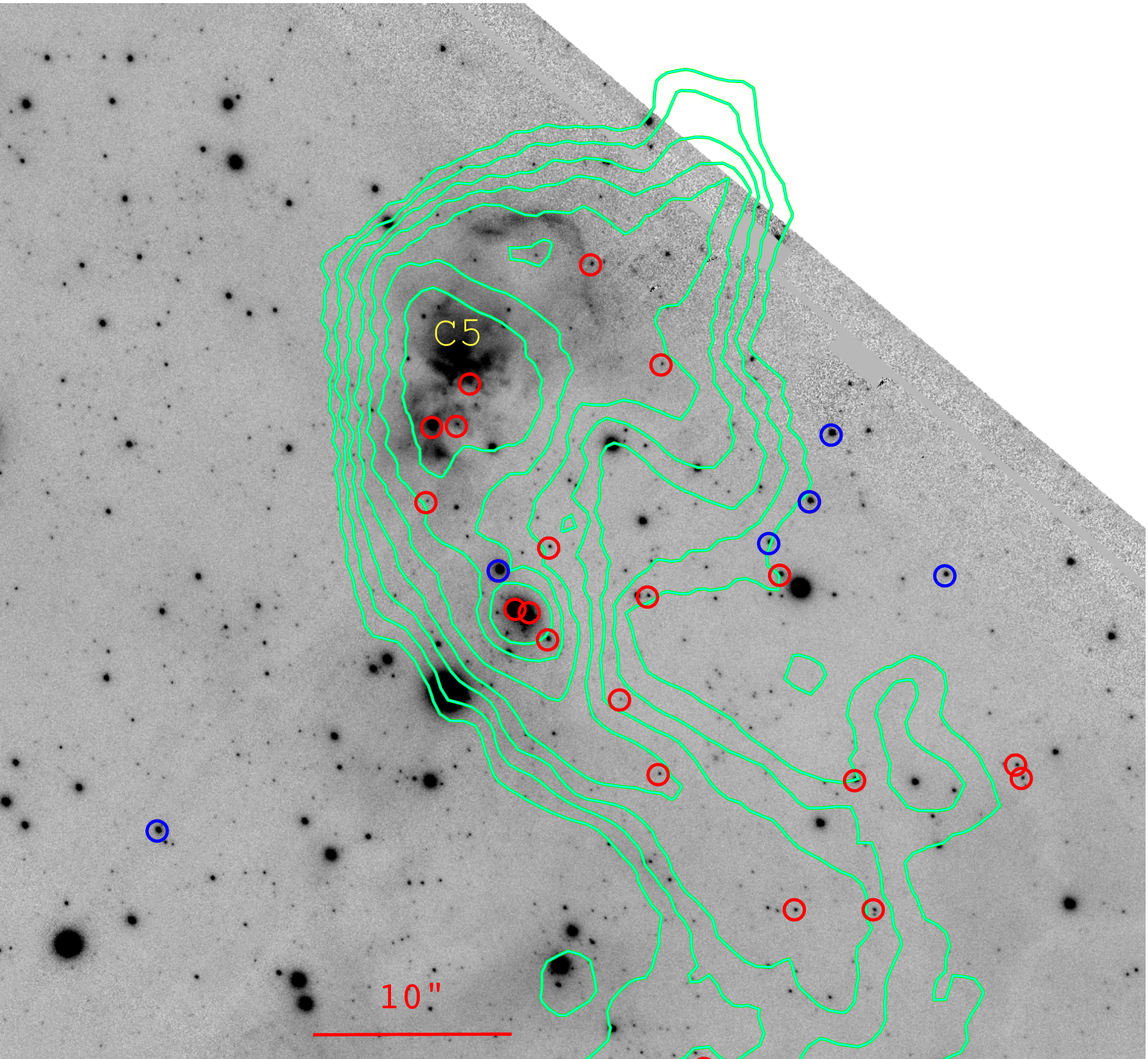} \\
\end{tabular}
\caption{Locations of the candidate YSOs in the GeMS/GSAOI \ks-band
  image, compared with the intensity of 8\,$\mu$m emission from {\em
    Spitzer} observations \citep[green contours,][]{Jones2005}; north
  is up and east is left. The blue circles show YSO candidates from
  analysis of the colour-colour diagram, red circles are those from
  analysis of the colour-magnitude diagram, see text for details. The
  left-hand panel shows the region around the central cluster
  (indicated by the white circle), with the yellow arc showing the
  location of an apparently limb-brightended rim; the right-hand panel
  shows the region around the C5 compact
  clump.}\label{fig:spatial-loc}
\end{figure*}

The spatial distribution of YSOs in a star-forming complex provides an
excellent tracer of recent star formation. The locations of our
candidate YSOs in the region of the central cluster and the C5 clump
are shown in the left- and right-hand panels of
Fig.~\ref{fig:spatial-loc}, respectively, including candidates from
analysis of the CCD (blue circles) and CMD (red circles). Also shown
are intensity contours of 8\,$\mu$m emission from the {\em Spitzer}
IRAC data \citep{Jones2005}. The locations of the candidate
YSOs are generally correlated with the 8\,$\mu$m emission, which also
corresponds to dense clumps of molecular gas
\citep[e.g.][]{Seale2012}. In contrast, there are only a few
candidate YSOs in the central cluster region (indicated by the white
circle in Fig.~\ref{fig:spatial-loc}), where the 8\,$\mu$m emission is
less intense.

The 8\,$\mu$m morphology around the cluster resembles many Galactic
bubbles associated with \hii regions \citep[e.g.][]{Deharveng2011}.
The 8\,$\mu$m IRAC band contains emission features at 7.7 and
8.6\,$\mu$m which are commonly attributed to polycyclic aromatic
hydrocarbons \citep[PAHs; e.g.][]{Peeters2004}.  PAHs are believed to
be destroyed in ionised gas \citep{Pavlyuchenkov2013}, but are thought
to be excited in the photo-dissociation region at the interface of a
\hii region and a molecular cloud, by the absorption of far-UV photons
from the exciting stars of the \hii region.  In particular, PAH
emission is thought to be a good tracer of the formation of B-type
stars \citep{Peeters2004}, which heat the surrounding dust to high
temperatures, exciting the PAH bands and fine-structure lines.
Indeed, bright, compact emission around B-type stars has been seen in
Galactic bubbles
\citep[e.g.][]{Zavagno2007,Samal2014,Dewangan2015,Kerton2015}.

In summary, the locations of the candidate YSOs support the idea that
the compact 8\,$\mu$m emission and molecular clumps in N159W appear to
be associated with recent massive-star formation, \citep[see
also][]{Jones2005,Testor2006}. In contrast, the relative absence of
YSOs in the central cluster suggests it was formed earlier. Possible
scenarios for the history of the region are discussed in
Sect.~\ref{sec:scenario}.


\section{Properties of the central cluster}

We now discuss the properties of the central cluster, defined as the
sources detected within the white circle in Figs.~\ref{fig:threecolor}
and \ref{fig:spatial-loc}.


\subsection{Age}\label{sect:age}

\begin{figure}
\centering
\includegraphics[trim=0cm 0cm 0cm 2cm,width=11.5cm,height=10.5cm]{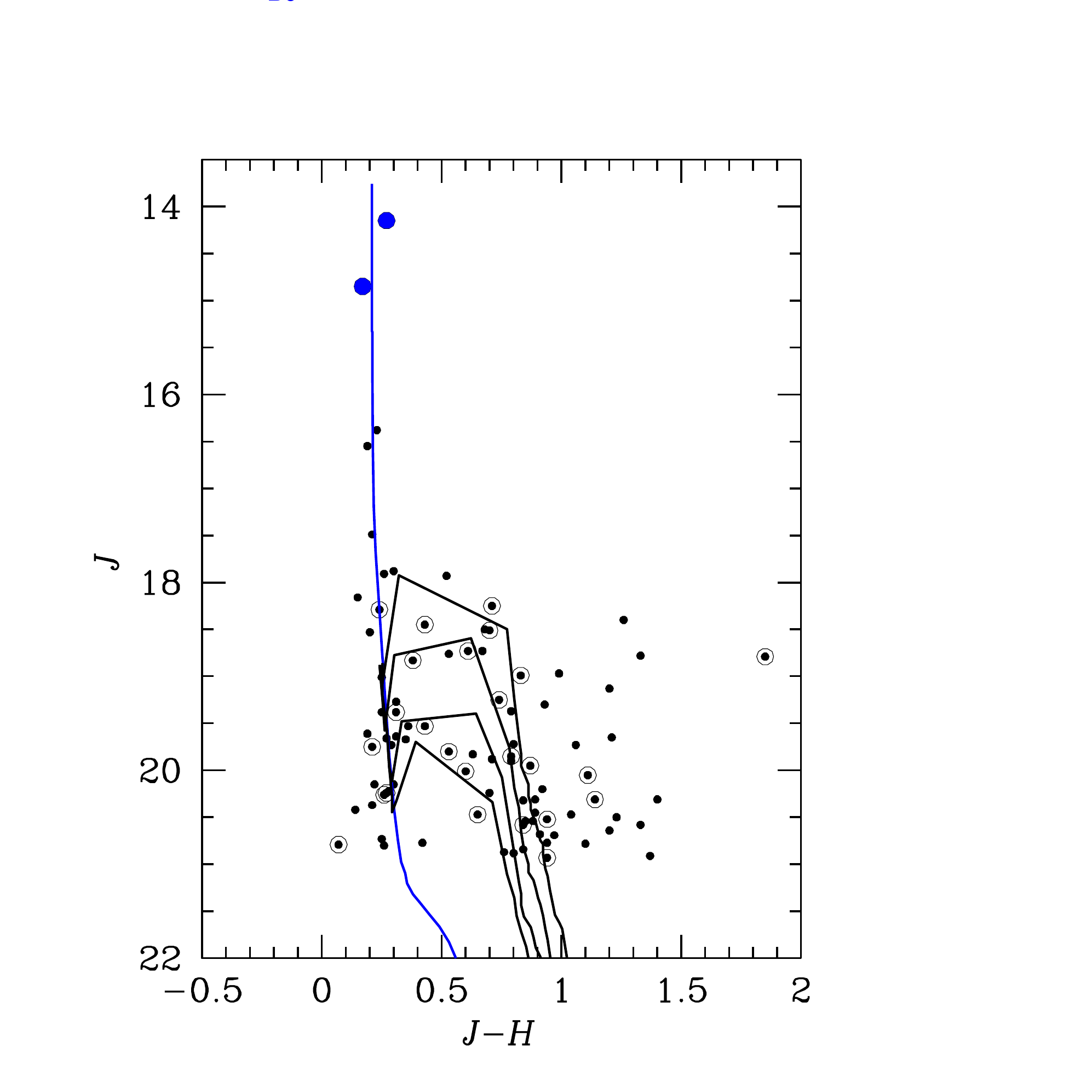}
\caption{Colour-magnitude diagram of stars in the region of the
  central cluster, with the two central O-type stars indicated by the
  large blue symbols. PMS isochrones from \citet[][for
  $Z$\,$=$\,0.01]{Siess2000} and ages of 0.5, 1, 2, and 3\,Myr (from
  right to left), corrected for a distance of 50\,kpc and average
  extinction of $A_V$\,$=$\,2.8\,mag, are indicated by the solid black
  lines, with the zero-age main sequence from \citet{Marigo2008} shown
  by the blue line.}\label{fig:JJH}
\end{figure}

We estimated the age of the central cluster using the CMD of its
member stars (see Fig.~\ref{fig:JJH}). We preferred to use the
$J$-band here rather than $H$ or \ks\/ to minimize the effect of
potential NIR-excesses on the age estimate. The zero-age main sequence
\citep[blue line,][]{Marigo2008} and PMS isochrones for ages of
0.5-3\,Myr \citep{Siess2000} are also plotted in the figure, after
correction for the distance (50\,kpc) and mean extinction
\citep[$A_V$\,$=$\,2.8\,mag,][]{Testor2006}.

The main sequence is well defined (albeit sparsely populated) down to
$J$\,$\sim$\,18.5\,mag in Fig.~\ref{fig:JJH}. At fainter magnitudes a
number of redder stars are seen and, from comparison with the
isochrones from \citet{Siess2000}, some are probably PMS stars. The
majority of these stars are located redwards of the 3\,Myr isochrone
(after correction for the mean extinction), indicating that the
cluster is unlikely to be older than 3\,Myr.  However, background
sources could also populate this PMS zone. To minimise such
contamination we selected only sources within a radius of 4$''$
(1\,pc) of the central massive stars, where the stellar density is
greatest. These sources are marked with open circles in
Fig.~\ref{fig:JJH}, and most are located between the 0.5 and 3\,Myr
isochrones, confirming the young age of the cluster. From this
spatially-restricted subsample the average age estimated for the
central cluster is 2\,$\pm$\,1\,Myr.

\subsection{Mass function}   

We also investigated the IMF and total mass of the central cluster. We
limited our analysis to the region indicated by the white circle in
Fig.~\ref{fig:threecolor} as it represents a well-bounded sample (in
terms of age, metallicity and distance), and the observed present-day
mass function should then be a fair representation of the underlying
IMF. Here we used the $H$-band photometry, as the number of sources
and photometric accuracy were better cf. the other bands (while also
avoiding potential NIR excesses which might influence the \ks-band
photometry).

\begin{figure}
   \centering
 \includegraphics[width=\hsize]{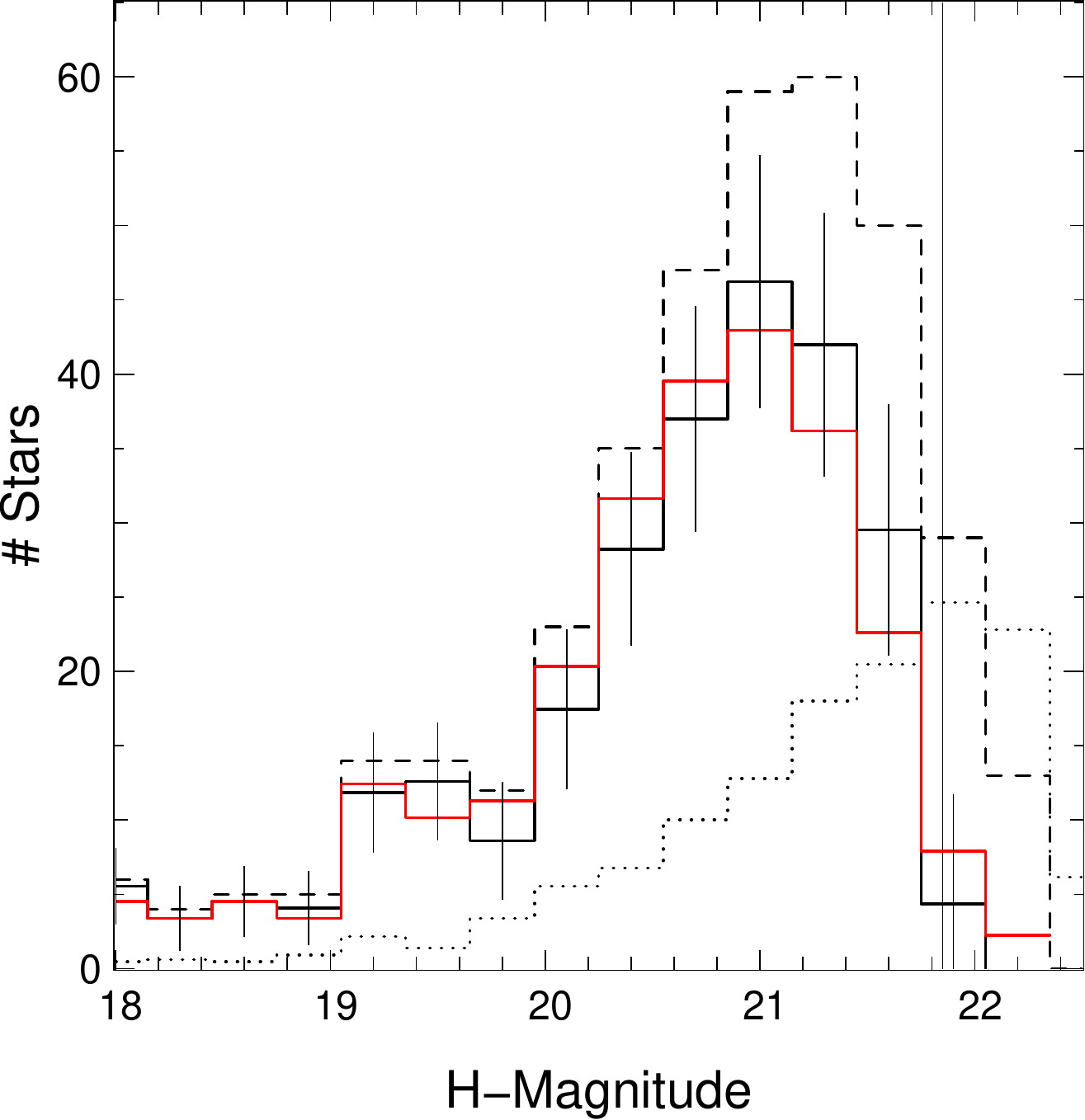}
 \caption{$H$-band luminosity functions for the central cluster
   (dashed line), and the control field (dotted line). The luminosity
   functions are also shown after correction for field-star
   contamination, using the control field (black solid line) and the
   colour-cut selection (red line). Error bars are estimated from
   Poisson statistics, and the (90\%) completeness limit is indicated
   by the vertical solid line ($H$\,$=$\,21.85\,mag).}\label{fig:HLF}
\end{figure}

In constructing the luminosity function for the cluster we corrected
for possible field-star contamination using two
approaches\footnote{Note that by adopting a completeness criterion of
  90\% in Sect.~\ref{subsec:comp}, the impact of incompleteness below
  this magnitude limit is negligible.}. The first method used
statistics from the control field, while the second attempted to
disentangle cluster and field stars based on their location in the
$H$\,vs.\,$H-\ks$ CMD \citep[see][]{Neichel2015a}. Corrections
were done for magnitude bin by subtracting the estimated number
of field stars from the detected source counts. The two methods
gave consistent results, as shown in Fig.~\ref{fig:HLF}, in which
estimated uncertainties are from Poisson statistics (i.e, the square
root of the number of stars). As our sample is limited in size, these
uncertainties are the primary limitation on determination of the
luminosity function.

\begin{figure}
   \centering
 \includegraphics[width=\hsize]{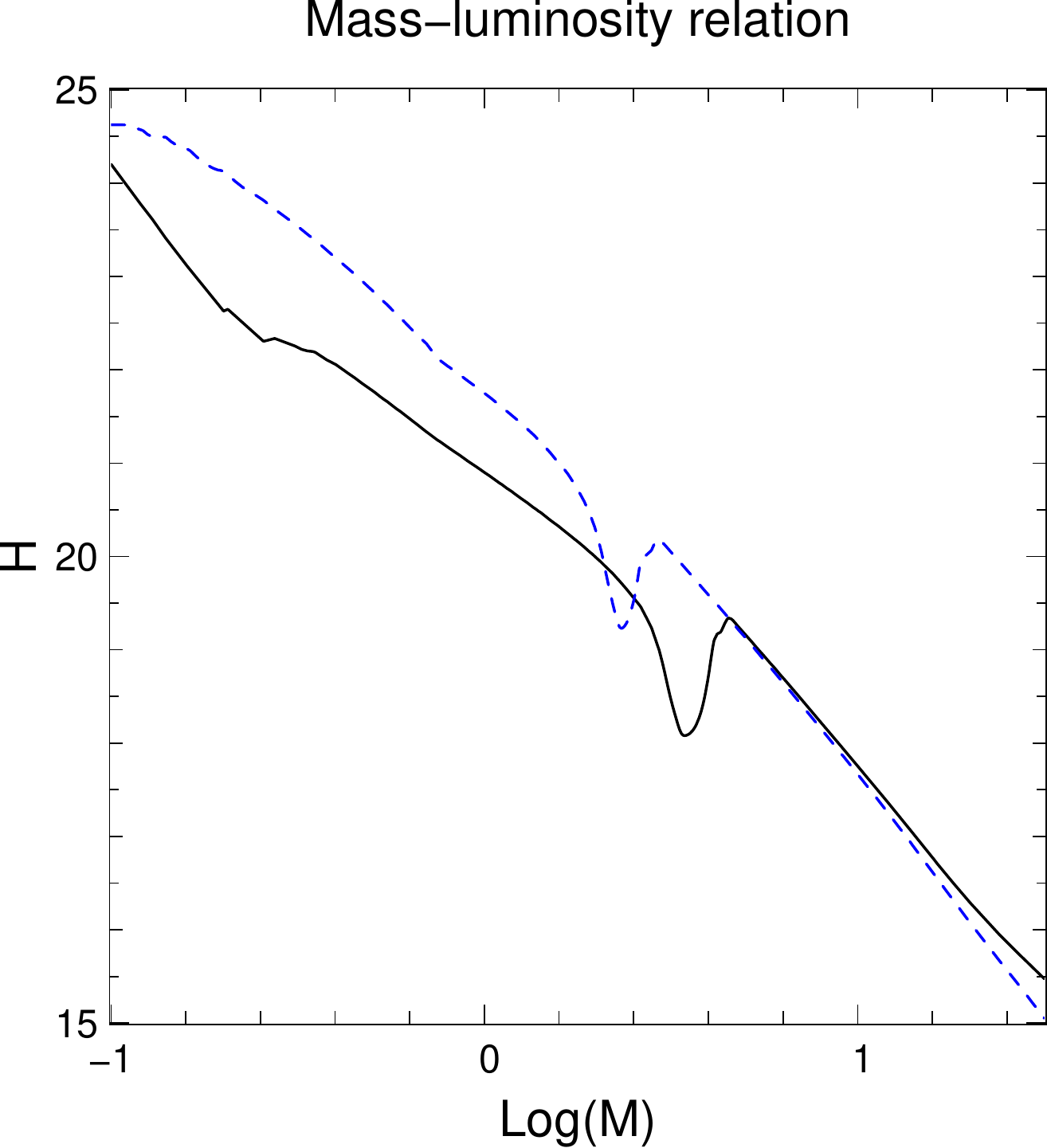}
 \caption{Mass-luminosity relation for the
   1\,Myr (black solid) and 3\,Myr (blue dashed)
   isochrones.}\label{fig:MLR}
\end{figure}

\begin{figure}
   \centering
 \includegraphics[width=\hsize]{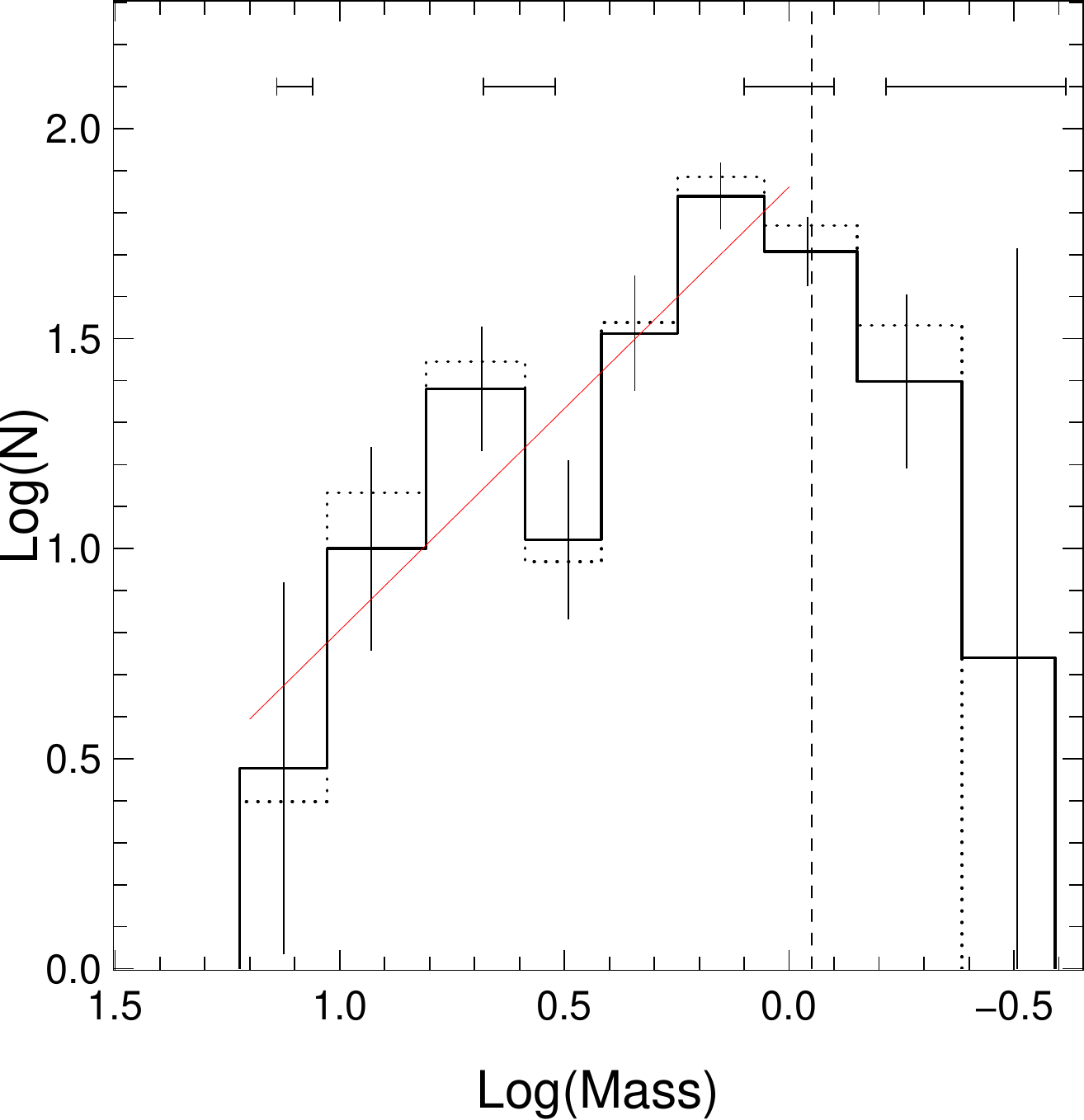}
 \caption{Present-day mass function (MF) of the central cluster of N159W.
The solid histogram is the MF computed after correction from field stars
using the control field; the dotted histogram shows the MF corrected
using the CMD analysis (see text for details). Uncertainties in the
mass estimates are shown as horizontal lines above the MF, with the
vertical lines in each bin indicating the Poisson uncertainties.
Fit to the MFs is shown in red, and is consistent with 
$\Gamma$\,$=$\,$-$1.05\,$\pm$\,0.2. The 90\% completeness limit
is shown by the dashed vertical line.}\label{fig:IMF}
\end{figure}

To construct the present-day mass function of the cluster we converted
the luminosity function using mass-luminosity relations
(Fig.~\ref{fig:MLR}), constructed from isochrones from
\citet{Siess2000} for $M$\,$<$\,6\,\msun, and from \citet{Marigo2008}
for more massive stars, adopting the same distance and average
extinction as before. Given the uncertainty in the exact age of the
cluster, we estimated the mass function using both the 1 and 3\,Myr
relations, and then calculated the average to account for a small
spread of ages. The final mass function, assumed to be representative
of the IMF given the young age, is shown in Fig.~\ref{fig:IMF}.  The
typical uncertainties on the mass estimates are indicated above the
mass function, and the estimated uncertainties on the number in each
bin are again from Poisson statistics.

We do not expect significant variations in NIR extinction across the
central cluster, so adopting a mean value in this analysis should have
a limited impact on the estimated masses. To investigate this further
we estimated the $\ks$-band extinction from the $H-\ks$ colours
\citep[as described by][]{Gutermuth2005}, finding an average $A_{K\rm
  s}$\,$=$\,0.3\,$\pm$\,0.05\,mag, with peak values between
0.2\,$<$\,$A_{K\rm s}$\,$<$\,0.4\,mag. This is consistent with the
estimated optical extinction ($A_V$\,$=$\,2.15 and 3.54\,mag) toward
the two central massive stars \citep{Testor2006}. Thus, the impact of
variable extinction is minimal compared to the potential age spread.

The IMF estimated from Fig.~\ref{fig:IMF} spans masses of $\sim$0.5-
10\,\msun, and shows the typical rise in number of stars with
decreasing mass into the subsolar regime. A simple fit over this mass
range gave a power-law slope of $\Gamma$\,$=$\,$-$1.05\,$\pm$\,0.2,
cf. the canonical value of $\Gamma$\,$=$\,$-$1.35
\citep{Salpeter1955}. We experimented with changing the bin sizes and
mass limits of the fit, but the results were consistent with the
quoted uncertainty of $\pm$\,0.2\,dex.

The IMF for the cluster in N159W appears to peak at a mass between 0.5
and 1\,\msun, although we are limited by our uncertainties in the mass
estimates in this range; incompleteness also limits any conclusions
for lower masses ($M$\,$<$\,0.5\,\msun) from our data. Nonetheless, we
note that such a flattening of the IMF at $\sim$0.5\,\msun\/ is
found in Galactic clusters \citep[e.g.][] {Kroupa2001, Kroupa2002,
  Luhman2007, Bastian2010, Dib2014} and in other clusters in the
Magellanic Clouds
\citep[e.g.][]{Dario2009,Andersen2009,Liu2009a,Gouliermis2012}.

Only deep {\em HST} observations can currently investigate the
lower-mass IMF in clusters in the Clouds. The only example to date is
the study of NGC\,1818 (with an age of 20-45\,Myr) by
\citet{Liu2009b}, who investigated the IMF down to 0.15\,\msun,
concluding that it was best described by a broken power-law
($\Gamma$\,$=$\,0.46\,$\pm$\,0.10 for 0.15\,$<$\,$M$\,$<$\,0.8\,\msun,
and a Salpeter-like slope, $\Gamma$\,$\sim$\,$-$1.35, at higher
masses) and that the estimated turn-over mass was consistent with the
0.5\,\msun\/ from \citet{Kroupa2001} given the uncertainties.

The N159W data discussed here are representative of the limitations of
AO-corrected observations from current 8-10\, class telescopes.
Robust investigation of the low-mass regime of the IMF in the Clouds
requires imaging with both the high sensitivity and angular resolution
provided by, e.g. the MICADO instrument \citep{Davies2010} under
construction for the 39\,m European Extremely Large Telescope.

When integrating over all the bins of the IMF, down to a mass limit of
0.9\,\msun, we estimated a total mass of 650\,$\pm$\,60\,\msun\/ for
the central cluster. If the cluster IMF indeed follows a canonical
power-law \citep{Kroupa2001,Kroupa2002,Kroupa2013}, we can infer that
around half of its mass is in the subsolar regime, giving a total
mass estimate of $\sim$1300\,\msun.

The mass of the most massive star in a cluster is generally found to
be linked to the total cluster mass \citep[see,
e.g.][]{Weidner2010b,Weidner2013,Popescu2014}, although there are
candidates of relatively isolated massive stars which challenge this
as a universal relationship \citep[e.g.][]{Bressert2012}. With the
earliest-type star in the cluster classified as O5-6, and assuming it
is a dwarf given the young age, at LMC metallicity this is expected to
have an initial mass in the range of 37-44\,\msun\/
\citep{Weidner2010a}. Although at the lower end, this is within the
expected range of the maximum mass from the simulations of
\citet{Popescu2014}, i.e. the maximum stellar mass in the N159W
cluster appears consistent with the cluster-mass relation.


\section{Star-formation scenario of the complex}
\label{sec:scenario}


Before investigating star-formation scenarios for the complex, we
investigated the dynamics of the compact clumps and central cluster.
ATCA observations from \citet{Seale2012} of HCO$^{+}$ (J = 1
$\rightarrow$ 0) and HCN (J = 1 $\rightarrow$ 0) revealed five
significant clumps of gas in the GeMS field: nos. 1, 7, 10, 12 and 15
in their Fig.~3 (with mass estimates in the range of
800-33\,000\,\msun), corresponding, respectively, to the C3, C5, C1,
C2, and C4 compact clumps identified here.  The radial velocities for
these clumps are in the range of 230-240\,km\,s$^{-1}$ (Fig.~6 from
\citeauthor{Seale2012}), in good agreement with the velocity of
ionised gas observed in the direction of the N159W cluster
\citep{Paron2016}, and confirming that they are dynamically
associated.

\cite{Seale2012} hypothesised that the massive clumps are sites of
cluster formation, and identified YSOs which appear associated.  Our
high angular-resolution images reveal that the clumps actually
comprise several YSOs, and that young cluster formation is indeed
underway at these locations. In contrast, we note that no HCO$^{+}$ or
HCN dense gas was observed by \cite{Seale2012} in the direction of
central cluster, indicating that star formation there is at a more
advanced stage. Indeed, from the locations of the PMS stars in
Fig.~\ref{fig:JJH}, we argued in Sect.~\ref{sect:age} that the central
cluster formed 1-3\,Myr ago.

Thus, a possible scenario is that the central O-type stars 
ionised the region around them which, when expanding, swept-up the ambient
gas and dust into a shell (accounting for the low level of 8\,$\mu$m
emission within the bubble). The distribution of the candidate YSOs in
close vicinity of the bubble suggests they may have been
triggered by its expansion, similar to star formation seen at the
edges of several geometrically-simple Galactic bubbles \citep[see,
e.g.][]{Deharveng2010a, Ohlendorf2013, Samal2014, Dewangan2015,
  Liu2015}. If this is really sequential star-formation in N159W, we
would expect the YSOs to be younger than the stars in the main
cluster. 

Unfortunately, from the current data we have no way to estimate the
ages of the embedded YSOs. Nonetheless, we searched for signatures of
the early phases of high-mass star formation such as masers
\citep[e.g. emission from water masers has a characteristic lifetime
of 2.5-4.5\,$\times$\.10$^4$\,yr;][]{vander2005}, UCHII regions
\citep[which typically only last for $\sim$3\,$\times$\,
10$^{5}$\,yr;][]{Mottram2011} and molecular outflows (generally
associated with type 0/I YSOs). The presence of outflows in C2 and C5
\citep{Fukui2015} and UCHII regions in C3 and C5 \citep{Hunt1994},
suggests star formation that is not older than a few
$\times$\,10$^5$\,yr.

If true, this favours an age difference between the central cluster
and the YSOs in the peripheral clumps, pointing to sequential star
formation. This appears particularly plausible for C1, which has an
associated limb-brightened rim in the \ks-band image (highlighted by
the yellow arc in Fig.~\ref{fig:spatial-loc}), with its apex pointing
toward the massive stars at the centre of the cluster. The morphology
of this structure resembles those from numerical simulations
\citep{Miao2009,Bisbas2011} and observations of bright-rimmed clouds
\citep[e.g.][]{Sugitani2002, Ikeda2008, Panwar2014}. The latter are
thought to arise from the impact of UV photons from nearby massive
stars on pre-existing dense molecular material
\citep{Lefloch1994,Miao2009}, leading to the formation of a new
generation of stars by radiatively-driven implosion. The morphology of
C1 suggests it is strongly influenced by ionising photons from the
central cluster, and the presence of candidate YSOs within the
clump provides support of a triggered scenario in this instance.

However, it is difficult to conclude such triggering is the only
formation process for the YSOs. Star-forming regions often contain
smaller cores/clumps with high column-densities, as revealed by {\em
  Herschel} observations \citep{Hill2011, Giannini2012, Hennemann2012,
  Schneider2012}, possibly due to density fluctuations present in the
original cloud. Provided the local conditions are close to
gravitationally bound, star formation can occur in these cores/clumps
on a local dynamical timescale. This can result in multiple stellar
groups, possibly in different evolutionary stages, in the same
complex, i.e. spontaneous star-formation can also occur close to a
\hii region.

For example, the cluster of YSOs detected in C2 coincides with the
N159W-S clump in the ALMA observations from \cite{Fukui2015}. As noted
earlier, they argued that the clump was formed by the collision of two
parsec-scale filaments.  If this is the case, then expansion of the
\hii region is not the prime mechanism in forming the clump, although
it may have brought the energy required to initiate star formation in
a clump formed via the filament collision. To further complicate
things, C2 also sits perfectly on the edge of the (1-2\,Myr old)
larger-scale bubble suggested by \citet{Jones2005} and
\citet{Nakajima2005} to have triggered star formation in N159E and
N159W, so its origins are even more ambiguous.
The C5 clump appears unrelated to the central cluster in N159W, and
may have been triggered by the larger-scale bubble. 
As in the case of C2, this may have been a pre-existing molecular
reservoir in which the pressure from expansion of the large-scale
bubble initiated star formation, although the current data to 
not allow firm conclusions.

Lastly, given the age of the central cluster (1-3\,Myr) and its
location outside of the larger-scale bubble, its formation is probably
not connected, providing support of the scenario from
\citet{Chen2010a} where star formation in N159W started spontaneously
(or at least, was not triggered by the larger bubble).


\section{Summary}

We have presented deep, high angular-resolution, NIR images of the
N159W region in the LMC. We have used these to explore the stellar
content of the central cluster, as well as the star-formation history
of the region. Based on these observations, via construction of
colour-colour and colour-magnitude diagrams, we have identified 104
candidate YSOs in the observed 90$''$\,$\times$\,90$''$ field
($\sim$22\,$\times$\,22\,pc). These candidates are prominently
distributed at the edges of the N159W \hii bubble, and are
concentrated into clumps and subclusters of stars, which are resolved
for the first time by the high angular-resolution of our observations.
These groups are associated with active star-formation processes, as
highlighted by the presence of UCHII regions, outflows, and class 0/I
YSOs -- each of which are indicators of very recent star-formation
events, possibly forming massive stars.

In contrast, the stars located inside the \hii bubble appear to be at
a more evolved stage, with an average age of 2\,$\pm$\,1\,Myr. The
central cluster has two massive O-type stars at its centre, which probably
dominate the ionisation of the \hii region. From the IMF of the
cluster, we estimate its total mass to be $\sim$1300\,\msun.  This
cluster may have instigated the star-formation events at its edges,
but further observations will be needed to firmly confirm the
presence of sequential star-formation in the N159W region.

\begin{acknowledgements}

  Based on observations obtained at the Gemini Observatory, which is
  operated by the Association of Universities for Research in
  Astronomy, Inc., under a cooperative agreement with the NSF on
  behalf of the Gemini partnership: the National Science Foundation
  (United States), the National Research Council (Canada), CONICYT
  (Chile), the Australian Research Council (Australia), Minist\'erio
  da Ci$\rm{\hat{e}}$ncia, Tecnologia e Inova{\c{c}}$\rm{\tilde{a}}$o
  (Brazil) and Ministerio de Ciencia, Tecnologiaa e Innovaci\'on
  Productiva (Argentine). B.~Neichel and A.~Bernard acknowledge the
  financial support from the French ANR program WASABI to carry out
  this work. H.~Plana thanks the CNPq/CAPES for its financial support
  using the PROCAD project 552236/2011-0. M.~R.~Samal acknowledges
  the financial support provided by the French Space Agency (CNES) for
  his postdoctoral fellowship.
\end{acknowledgements}


\bibliographystyle{aa}

\bibliography{library_ab.bib}

\end{document}